\documentclass{jfp}
\usepackage[utf8]{inputenc}
\usepackage[T1]{fontenc}
\usepackage{microtype}
\usepackage{graphicx}
\usepackage{xcolor}
\usepackage{caption}
\usepackage{subcaption}
\usepackage{epigraph}
\usepackage{natbib}
\usepackage[inline,short labels]{enumitem}
\usepackage{listings}
\usepackage{cleveref}

\makeatletter
\let\JFP@linecountL\relax
\makeatother

\renewcommand{\cite}{\citep}

\theoremstyle{theorem}
\newtheorem{theorem}{Theorem}[section]

\newtheorem{property}[theorem]{Property}
\newtheorem{fact}[theorem]{Fact}
\theoremstyle{definition}

\theoremstyle{remark}
\newtheorem{remark}{Remark}[section]

\crefname{fact}{fact}{facts}
\Crefname{fact}{Fact}{Facts}

\lstdefinelanguage{Agda}{
  language=Haskell,
  keywords={%
    module,open,%
    where,with,%
    data,inductive,coinductive,%
    record,constructor,field,%
  },
}
\lstdefinelanguage{Scheme}{
  morecomment=[l]{;},
  morecomment=[s]{/*}{*/},
  morestring=[b]",
  keywords={%
    define,define-syntax,syntax-rules,define-syntax-rules,%
    let,let*,letrec,%
    lambda,apply,eval,%
    cond,if,else,%
    equal?,eqv?,eq?,%
    quote,unquote,quasiquote,%
    define-record-type,fields,%
    mutable,immutable,%
    call/cc,with-exception-handler,%
    match,match*,match**,%
    match-lambda,match-lambda*,match-lambda**,%
    cocase,cocase*,comatch,%
  }
}
\newcommand{\ie}{{\em i.e., \/}}

\newcommand{\eg}{{\em e.g., \/}}

\newcommand{\aka}{{\em a.k.a \/}}

\newenvironment{contiguous}[1][\linewidth]
{\par\noindent\minipage{#1}}
{\endminipage\par\noindent\ignorespacesafterend}

\newcommand{\paul}[1]{}%{{\color{purple}\emph{Paul: #1}}}

\begin{document}

\journaltitle{arXiv}
\cpr{}
\doival{}
\def\doitext{}

\lefttitle{P.~Downen and Z.M.~Ariola}
\righttitle{Classical (Co)Recursion: Programming}

\totalpg{\pageref{lastpage01}}
\jnlDoiYr{2021}

\title{Classical (Co)Recursion: Programming}

\begin{authgrp}
\author{Paul Downen} and~ \author{Zena M. Ariola}
\affiliation{University of Oregon \\
        (\email{\{pdownen,ariola\}@cs.uoregon.edu})}
\end{authgrp}

%\received{1 March 2021}

\begin{abstract}
  Structural recursion is a widespread technique.  It is suitable to be used by
  programmers in all modern programming languages, and even taught to beginning
  computer science students.  What, then, of its dual: \emph{structural
    co\-recursion}?  For years, structural co\-recursion has proved to be an
  elegant programming technique made possible by languages like Haskell.  There,
  its benefits are to enable compositional algorithm design by decoupling the
  generation and consumption of (potentially) infinite or large collections of
  data.  However, it is usually thought of as a more advanced topic than
  structural recursion not suitable for beginners, and not easily applicable
  outside of the relatively narrow context of lazy, pure functional programming.

  Our aim here is to illustrate how the benefits of structural co\-recursion can
  be found in a broader swath of the programming landscape than previously
  thought.  Beginning from a tutorial on structural co\-recursion in the total,
  pure functional language Agda, we show how these same ideas are mapped to
  familiar concepts in a variety of different languages.  We show how
  co\-recursion can be done in strict functional languages like Scheme, and even
  escapes the functional paradigm entirely, showing up in the natural expression
  of common object-oriented features found in languages like Python and Java.
  Opening up structural co\-recursion to a much wider selection of languages and
  paradigms---and therefore, also to a much larger audience of
  programmers---lets us also ask how co\-recursion interacts with computational
  effects.  Of note, we demonstrate that combining structural co\-recursion with
  effects can increase its expressive power.  We show a \emph{classical} version
  of co\-recursion---using first-class control made possible by Scheme's
  classical \texttt{call/cc}---that enables us to write some new
  stream-processing algorithms that aren't possible in effect-free languages.
\end{abstract}

\maketitle

\section{Introduction}
\label{sec:introduction}

\setlength\epigraphwidth{0.45\textwidth}
\epigraph
{
 In a sense, recursive equations are the `assembly language' of
 functional programming, and direct recursion the \texttt{goto}.
}
{\cite{OrigamiProgramming}}

Recursion on the structure of recursive data types is a common principle for
designing practical programs.  This notion---based on the idea of induction on
the natural numbers in mathematics---has been fruitfully applied to programming
for decades, and is so well-understood that every year it is taught to swaths of
beginning computer science students as a general-purpose technique of algorithm
design \cite{HTDP}.  The advantage of structural recursion is to provide a
template to processing input data of any size by combining results given for
smaller sub-parts of that input, but with the guarantee that this process will
always finish with a final answer.  For example, the most basic form of
structural recursion is ``primitive recursion'' on numbers---corresponding to
``primitive induction'' and which we refer to as just ``recursion'' for
short---stipulates that recursive calls are only allowed on the immediate
predecessor of the input.  This differs from ``general recursion,'' which
imposes no restrictions at all on the recursive calls, and as such does not come
with any guidance for program design or guarantees of termination.  Moreover,
the technique of structural recursion extends far beyond just numbers, and can
capture algorithms over any inductively-defined data type, from lists to finite
trees of nearly any shape imaginable.

What, then, of the natural dual of structural recursion: \emph{structural
  co\-recursion}?  Co\-recursion has been used in several applications of
co\-algebras \cite{JacobsRutten97Tutorial,RuttenMethodofCoalgebra}.  But this is
quite different from the way structural recursion is presented, in its own
right, as an independent technique for practical program design.  And it is
certainly not a topic that is readily taught to beginning computer science
students in this form.  We believe this is a sadly missed opportunity, and
what's missing is a purely computational point of view for co\-recursion.
Whereas the structure of recursion directly follows the structure of a program's
\emph{inputs}, the structure of co\-recursion directly follows the structure of
a program's \emph{outputs} \cite{HTDCoPrograms}.  With this point of view, the
practice of co\-recursion can and should be taught to first-year students, right
alongside other fundamental methods of designing and structuring programs. Here,
we aim to further this goal by demystifying structural co\-recursion, making it
more suitable for widespread use in programming environments
\cite{GordonWhatIsCoinduction} that share a common hidden notion of
\emph{co\-data} \cite{CodataInAction}.

The primary goal of this paper is to introduce the basic principles of
primitive, structural co\-recursion in a variety of real programming languages.
The secondary goal of this paper is to point out the expressive power of
different notions of structural co\-recursion.  There are many different
formulations of structural recursion that have their own tradeoffs, creating an
impact on issues like ease of use, expressive power, and computational
complexity.  For example, the recursion scheme corresponding to primitive
recursion is sometimes called a ``paramorphism'' \cite{Meertens92}, with
``catamorphism'' \cite{Hinze13} sometimes used for plain iteration.  Identifying
and studying these various recursion schemes opens a world of laws and theorems
which can be applied by a compiler, such as the catamorphism fusion laws which
eliminate intermediate structures \cite{Malcom90}.  Likewise, there are
different formulations of structural co\-recursions with similar tradeoffs.
Primitive co\-iteration, which unfolds from a given seed, is called
``anamorphism'' \cite{MFP91}; primitive co\-recursion which can stop the
computation at any time is called ``apomorphism''
\cite{Vene98functionalprogramming}.  These notions of co\-recursion are all
usually studied in the context of pure functional languages like Haskell
\cite{Unfold99,Graham99,ProofMethodsCorecursive}, where the types of infinite
versus finite objects are conflated.  What happens when we need to distinguish
infinite streams from finite lists?  Or when computational effects are added?
Or when we move away from the functional paradigm entirely?  Is there any
difference to the expressive power of the various co\-recursion schemes, and if
so, what new algorithms do they enable?

To illustrate the diversity and expressiveness of co\-recursion in practice, we
present a series of examples in a variety of programming languages, which
feature different levels of built-in support for co\-recursion, co\-data,
computational effects, and programming paradigms.
We consider four different programming languages: Agda, Scheme, Python, and
Java.  These four languages let us illustrate the impact of how co\-recursion
appears depending on combinations of these different choices:
\begin{itemize}
\item \emph{Paradigm}: either functional (Agda and Scheme) or object-oriented
  (Python and Java).
\item \emph{Typing discipline}: either static (Agda and Java) or dynamic (Scheme
  and Python)
\item \emph{Computational Effects}: which could include either first-class
  control (Scheme) or exceptions and handlers (Python and Java), or nothing at
  all (Agda).
\end{itemize}

% {\color{red} 
% So here we isolate some particular patterns of recursion of interest in order to
% make them concrete.  We present examples in Agda---a dependently-typed, total,
% functional programming language---which lets us define new inductive data types,
% and define (terminating) recursive functions by pattern-matching over values of
% those types.  This lets us define some \emph{combinators} that abstract out the
% recursion from some examples of common functions.
% }
% \zena{we should delete this paragraph}

We begin in \cref{sec:agda-rec} with a review of primitive, structural recursion
as found in Agda---a dependently-typed, total, functional programming
language---where we isolate some common recursive patterns that can be
abstracted out in terms of concrete \emph{combinators}.  Following in
\cref{sec:agda-corec}, we dualize these recursive patterns in Agda to find
co\-recursive patterns. Rather than inspecting the structure of input via pattern
matching, we can inspect the structure of output via \emph{co\-pattern matching}
\cite{Copatterns}.
% Like recursion versus iteration, we compare co\-recursion versus
% co\-iteration: co\-recursion can be more efficient than co\-iteration by
% letting co\-recursive processes stop early.
Next, we consider in \cref{sec:scheme-corec} how co\-recursion can be expressed
in Scheme without built-in support for co\-patterns.  Despite the lack of
copatterns, Scheme introduces a distinct advantage: programmable, first-class
continuations made available by the \texttt{call/cc} control operator.
First-class control lets us capture a notion of \emph{classical co\-recursion}
which is more expressive than before.  For example, in
\cref{sec:classical-expressiveness} we show a stream algorithm that cannot be
expressed with the pure co\-recursive combinators like anamorphisms and
apomorphisms.

From there, we shift our attention away from the functional programming paradigm
over to the object-oriented one.  In \cref{sec:python-corec}, we illustrate how
the same patterns of structural co\-recursion can be expressed in Python---a
dynamically typed, object-oriented language.  This involves the use of
\emph{persistent} objects whose behavior does not change over time, in contrast
to the commonly-used \emph{ephemeral} objects where each method call might
affect its future behavior.  There, we analyze the use of exceptions to model
different forms of streams, adding new features like the ability to end early or
skip intermediate elements.  We conclude, in \cref{sec:java-corec}, by
discussing how static types in an object-oriented language like Java interact
with structural co\-recursion and its combinators.  In particular, subtyping
between static interfaces formalizes the hierarchy streaming types based on
their features, and allows us to distinguish between \emph{safe} operations---in
the sense that they will always produce some productive result in finite
time---versus \emph{unsafe} ones that might cause the program to hang.  To
conclude (\cref{sec:conclusion}), we hope this paper demonstrates how the general
design patterns for co\-recursion---describing how boundless data should be
generated, to be used elsewhere---are applicable to different programming
paradigms, too.

% We then continue in \cref{sec:corec-programming} with examples of programming
% with co\-recursion in a variety of programming languages using different
% paradigms (functional and object oriented), typing disciplines (static and
% dynamic), and computational effects (exceptions, first-class control, and
% none).

%%% Local Variables:
%%% mode: latex
%%% TeX-master: "corec-prog"
%%% End:

%\section{Recursion in Agda: Typed Inductive Patterns}
%\section{Agda: Inductive Patterns and Coinductive Patterns}
\section{Agda: Inductive Patterns}
\label{sec:agda-rec}

\lstset{language=Agda}

Like most statically typed functional languages, Agda lets programmers define
new types through data type declarations.  For example, a representation of the
Peano numbers are captured by this data type definition:
\begin{contiguous}
\begin{lstlisting}[language=Agda]
  data Nat : Set where
    zero : Nat
    succ : Nat → Nat
\end{lstlisting}
\end{contiguous}
\lstinline|Nat| is a type (\ie a \lstinline|Set|) defined \emph{inductively}
over the listed constructors: \lstinline|zero| represents the number 0 and
\lstinline|succ| returns the successor of any number.  The inductive property of
\lstinline|Nat| means all its values are uniquely built up from \emph{finite}
applications of \lstinline|zero| and \lstinline|succ|.\footnote{Any natural
  number $n$ is represented as $\mathtt{(succ^n~zero)}$, \eg 3 is
  $\mathtt{(succ(succ(succ~zero)))}$.}
%1 is represented as $\mathtt{(succ~zero)}$, 2 as
%$\mathtt{(succ~(succ~zero))}$, \etc so that any natural number $n$ is
%$\mathtt{(succ^n~zero)}$.}.

The inductive nature of \lstinline|Nat| is an essential tool for defining
operations over natural numbers.  First, the fact that the constructors
\lstinline|zero| and \lstinline|succ| create distinct values means that we can
\emph{pattern-match} over them, to inspect values and learn how they were
constructed.  Second, the fact that these values are finitely constructed means
it is well-founded to define functions by \emph{structural recursion}, by
referring to solutions to that same function on ``smaller'' arguments, where
``smaller'' for \lstinline|Nat| means the argument of \lstinline|succ|, \ie its
predecessor.  For example, the function for addition (\lstinline|plus m n|) can
be written by pattern-matching on the first argument in the following
structurally recursive way:
\begin{contiguous}
\begin{lstlisting}[language=Agda]
  plus : Nat → Nat → Nat
  plus  zero    n = n
  plus (succ m) n = succ (plus m n)
\end{lstlisting}
\end{contiguous}
Since Agda is a total programming language, it checks that \lstinline|plus|
terminates for any two \lstinline|Nat| arguments, and finds that it does because
the recursive call to \lstinline|plus m n| in the second clause is ``smaller''
than the original call \lstinline|plus (succ m) n|, since \lstinline|m| is
smaller than \lstinline|succ m|.

Since this pattern of well-founded structural recursion over \lstinline|Nat| is
so common, we could name it and use it many times.  If we abstract over the
particulars of addition, we arrive at a pattern of recursion that
\lstinline|iter|ates over a single \lstinline|Nat|ural number, providing only
the recursive result in the case of a \lstinline|succ|essor.  This is also known
as a catamorphism \cite{Hinze13}, and is defined as follows (the parameter in
\lstinline|{ }| is an implicit type parameter):
\begin{contiguous}
\begin{lstlisting}[language=Agda]
  iter : {A : Set} → Nat → A → (A → A) → A
  iter  zero    z s = z
  iter (succ n) z s = s (iter n z s)
\end{lstlisting}
\end{contiguous}
In addition to the main number being recursed over, \lstinline|iter| takes one
argument (\lstinline|z|) as the value to give in the base case and a function
(\lstinline|s|) to apply in the recursive case.  Because this iteration might be
used to produce any type of result, the return type is an arbitrary type
\lstinline|A|, so that the base case is \lstinline|z : A| and the recursive step
is \lstinline|s : A → A|.  With this abstraction in hand, we can give an equivalent
definition for addition in one line by instantiating \lstinline|iter|:
\begin{contiguous}
\begin{lstlisting}[language=Agda]
  plus' : Nat → Nat → Nat
  plus' m n = iter m n succ
\end{lstlisting}
\end{contiguous}
Likewise, we can define other operations like multiplication both manually (by
structural recursion) or as an application of \lstinline|iter| as follows:
\begin{contiguous}
\begin{lstlisting}[language=Agda]
  times : Nat → Nat → Nat
  times  zero    n = zero
  times (succ m) n = plus n (times m n)
\end{lstlisting}
\end{contiguous}
\begin{contiguous}
\begin{lstlisting}[language=Agda]
  times' : Nat → Nat → Nat
  times' m n = iter m zero (plus n)  
\end{lstlisting}
\end{contiguous}

However, some functions resist fitting into the \lstinline|iter| mold.  For
example, the predecessor function is the inverse of \lstinline|succ|.  It can be
easily defined by pattern-matching as:
\begin{contiguous}
\begin{lstlisting}[language=Agda]
  pred : Nat → Nat
  pred  zero    = zero
  pred (succ n) = n
\end{lstlisting}
\end{contiguous}
where we take the predecessor of \lstinline|zero| to be again \lstinline|zero|.
There does not seem to be a direct
way to recover this definition by applying \lstinline|iter|.  Instead, we can
define another abstraction which captures this form of shallow \emph{case
  analysis} that does not recurse, and use it to define the predecessor
function, like so:
\begin{contiguous}
\begin{lstlisting}[language=Agda]
  case : {A : Set} → Nat → A → (Nat → A) → A
  case  zero    z s = z
  case (succ n) z s = s n
\end{lstlisting}
\end{contiguous}
\begin{contiguous}
\begin{lstlisting}[language=Agda]
  pred' : Nat → Nat
  pred' n = case n zero (λ n → n)
\end{lstlisting}
\end{contiguous}

% \begin{lstlisting}[language=Agda]
%   minus : Nat → Nat → Nat
%   minus m  zero    = m
%   minus m (succ n) = pred (minus m n)

%   minus' : Nat → Nat → Nat
%   minus' m n = iter n m pred
% \end{lstlisting}

But what happens if we need to do both recursion and shallow pattern matching at
the same time?  For example, consider the well-known factorial function, defined
as:
\begin{contiguous}
\begin{lstlisting}[language=Agda]
  fact : Nat → Nat
  fact  zero    = succ zero
  fact (succ n) = times (succ n) (fact n)
\end{lstlisting}
\end{contiguous}
The definition of \lstinline|fact (succ n)| refers to the recursive call
\lstinline|(fact n)| \emph{and also} to the number \lstinline|n| itself .  This
pattern of structural recursion, also known as paramorphism \cite{Meertens92},
does not seem to directly fit either \lstinline|iter| or \lstinline|case|.
Instead, here is an even more general form of primitive \lstinline|rec|ursion
which performs both tasks simultaneously:
\begin{contiguous}
\begin{lstlisting}[language=Agda]
  rec : {A : Set} → Nat → A → (Nat → A → A) → A
  rec  zero    z s = z
  rec (succ n) z s = s n (rec n z s)
\end{lstlisting}
\end{contiguous}
In \lstinline|rec|, the function \lstinline|s| applied in the recursive step for
\lstinline|succ n| is provided \emph{both} the predecessor \lstinline|n| as well
as the recursive result \lstinline|(rec n z s)|.  Supplying both pieces of
information makes it straightforward to redefine \lstinline|fact| in terms of
\lstinline|rec| like so:
\begin{contiguous}
\begin{lstlisting}[language=Agda]
  fact' : Nat → Nat
  fact' m = rec m (succ zero) (λ n x → times (succ n) x)
\end{lstlisting}
\end{contiguous}

% \begin{remark}
% Iteration is straightforward to express in terms of recursion like so:
% \begin{lstlisting}[language=Agda]
%   iter  n z s =  rec n z (λ _ x →  s x)
% \end{lstlisting}
% Though it may not be obvious, the reverse encoding is also possible, though at a
% serious cost.  The encoding makes use of pairs and is expressed in Agda as:
% %\lstinline|rec| can be redefined in terms of \lstinline|iter| in Agda as:
% \begin{contiguous}
% \begin{lstlisting}[language=Agda]
%   id-and-rec' : {A : Set} → Nat → A → (Nat → A → A) → Nat × A
%   id-and-rec' n z s = iter n (zero , z) λ{(x , y) → (succ x , s x y)}
% \end{lstlisting}
% \end{contiguous}
% \begin{contiguous}
% \begin{lstlisting}[language=Agda]
%   rec' : {A : Set} → Nat → A → (Nat → A → A) → A
%   rec' n z s = proj₂ (id-and-rec' n z s)
% \end{lstlisting}
% \end{contiguous}
% The helper function \lstinline|id-and-rec'| recomputes the given number
% \lstinline|n| (the first component of the returned pair) simultaneously while
% computing the actual recursive result of interest (the second component).  Doing
% so lets us access both pieces of information in the iterative step, which get
% bound to \lstinline|(x , y)| by the $\lambda$-abstraction.
% % More details of this encoding
% %  will be discussed later in \cref{sec:rec-vs-iter}.
% \end{remark}

%\zena{Refer to Wadler's book?}

% \zena{I would delete the stuff about high-order recursion: max, max',min,min'.
% Or otherwise leave max, max' and delete min and min'. Same for the lists, only give maximum}

We have covered recursion over one \lstinline|Nat|, but what about two?  For
example, consider this definition for calculating the maximum of two natural
numbers:
\begin{contiguous}
\begin{lstlisting}[language=Agda]
  max : Nat → Nat → Nat
  max  zero     n       = n
  max (succ m)  zero    = succ m
  max (succ m) (succ n) = succ (max m n)
\end{lstlisting}
\end{contiguous}
\lstinline|max| pattern-matches against \emph{both} of its arguments in the
latter two clauses, and in the case where they are both \lstinline|succ|essors,
it recurses on the predecessor of the two.  Is this definition well-founded?
Yes, it is still an instance of structural recursion because (both) arguments to
the recursive call of \lstinline|max| are smaller.  But how can \lstinline|max|
fit into the pattern of \lstinline|rec|, which only iterates on a single number?
We could select one of the arguments as the ``main'' number for
\lstinline|rec|ursion---in \lstinline|max|, the first argument seems to fit well
since every clause matches on it.  Then the other argument can be inspected by
shallow \lstinline|case| analysis, as necessary.

However, there is still an issue: even if the first argument of \lstinline|max|
is the main one for purposes of recursion, \emph{both} arguments still change in
the recursive call.  Thankfully, there is another trick we can use:
\lstinline|rec| can compute \emph{any} type of result.  Previously, all our
applications of \lstinline|rec|, \lstinline|iter|, and \lstinline|case| (used to
define \lstinline|plus'|, \lstinline|times'|, \lstinline|pred'|, and
\lstinline|fact'|) calculated just a single \lstinline|Nat| number.  Here, we
can instead use \lstinline|rec| to calculate a \emph{function} of type
\lstinline|Nat → Nat| which corresponds to the partial application
\lstinline|max m|.  In other words, we can use \lstinline|rec| to perform
\emph{higher-order recursion} \cite{HarperPFPL,Turner}.  This function, in turn,
accepts the second argument to \lstinline|max| and may inspect that number to
respond appropriately.  The alternative definition of \lstinline|max| can be
given in terms of \lstinline|rec| as follows, where we annotate the return type
of \lstinline|rec| and \lstinline|case| by providing the implicit type parameter
(in braces \lstinline|{ }|):
\begin{contiguous}
\begin{lstlisting}[language=Agda]
  max' : Nat → Nat → Nat
  max' m = rec {Nat → Nat} m
    (λ n → n)
    (λ m' f n → case {Nat} n
      (succ m')
      (λ n' → succ (f n')))  
\end{lstlisting}
\end{contiguous}
In the latter two arguments of \lstinline|rec|, the $\lambda$-bound
\lstinline|n| corresponds to the second argument to \lstinline|max'|.
Additionally, the $\lambda$-bound \lstinline|m'| stands for the predecessor of
the first argument \lstinline|m|, and \lstinline|f| stands for the partial
application of \lstinline|max' m'|.  Likewise, the $\lambda$-bound
\lstinline|n'| in the last argument of \lstinline|case| stands for the
predecessor of the second argument \lstinline|n|.  Notice how the three possible
responses---\lstinline|n|, \lstinline|(succ m')|, and \lstinline|succ (f n')|%
---correspond one-for-one with the right-hand sides to the three clauses
defining \lstinline|max| above.  In particular, \lstinline|succ (f n')| is
equivalent to \lstinline|succ (max' m' n')|, which is the same as the third
clause defining \lstinline|max|.

\section{Agda: Typed Coinductive Copatterns}
\label{sec:agda-corec}

\lstset{language=Agda}

As a prototypical example of a coinductive type, consider the type of infinite
streams, which is defined like so in Agda:
\begin{contiguous}
\begin{lstlisting}[language=Agda]
  record Stream A : Set where
    coinductive
    field head : A
          tail : Stream A
\end{lstlisting}
\end{contiguous}
Instead of a data type with multiple constructors, \lstinline`Stream A` is
defined as a \lstinline`record`: a form of collection which contains a number of
\lstinline`field`s.%
\footnote{ \lstset{basicstyle=\ttfamily\footnotesize} This record can be seen as
  a recursively defined \emph{codata type} \cite{CodataInAction}, where the
  field accessors are the destructors.  All operations on \lstinline`Stream A`
  objects boil down to a combination of \lstinline`head` and \lstinline`tail`.}
In this case, the two fields of a \lstinline`Stream A` are
a \lstinline`head` of type \lstinline`A` and a \lstinline`tail` of type
\lstinline`Stream A`.  Because of the recursion in the type (the
\lstinline`tail` of a stream is another stream), we have to declare that this
type is \lstinline`coinductive` to say \lstinline`Stream A` objects are infinite
(whereas \lstinline`inductive` only allows for finite objects).

% Lastly, this declaration names a constructor for streams: the binary operator
% \lstinline`x ,, xs` that combines a head \lstinline`x` and tail \lstinline`xs`
% into a new stream.  This constructor might sound similar to $\<scons>$ from
% \cref{ex:scons}, and in fact one can be defined in terms of the other, like
% so:
% \begin{contiguous}
% \begin{lstlisting}[language=Agda]
%   scons : {A : Set} → A → Stream A → Stream A
%   scons x xs = x ,, xs
% \end{lstlisting}
% \end{contiguous}

Since \lstinline`head` and \lstinline`tail` only provide ways of using
\lstinline`Stream`s, how do we create them?  Agda uses \emph{co\-patterns}
\cite{Copatterns} as a particularly useful way for creating values of
coinductive types like \lstinline`Stream`.
%Agda extends the traditional
%functional style of defining functions by pattern-matching \emph{co\-patterns}
%\cite{Copatterns}, which are particularly useful for creating values of
%coinductive types like \lstinline`Stream`.
For example, when defining the \lstinline`plus` function in \cref{sec:agda-rec},
we gave different answers based on the shape of its arguments.  So rather than
just giving one definition for a generic call \lstinline`plus m n`, we answered
two different questions: what is \lstinline`plus zero n` equal to? (it's
\lstinline`n`) and what is \lstinline`plus (succ m) n` equal to? (it's
\lstinline`succ (plus m n)`).  These two equalities hold \emph{by definition},
and it is up to the implementation to figure out the details of branching checks
and jumps that satisfy them.

Co\-patterns use this same idea of giving multiple equations, but also allow the
programmer to match on the shape of the \emph{projections} in the context of a
definition, not just the arguments in the calling context.  In other words,
co\-patterns are a method of designing a program based on the shape of its
\emph{output} \cite{HTDCoPrograms}, complementing patterns that consider the
shape of \emph{input}.  Consider how to describe a stream built by adding a new
element on the front.  Lets write this stream as \lstinline`x ,, xs`, where the
single element \lstinline`x` is followed by the infinite stream \lstinline`xs`.
The two primary questions to answer about this stream correspond exactly to the
field projections: what is \lstinline`head (x ,, xs)` equal to? (it's
\lstinline`x`) and what is \lstinline`tail (x ,, xs)` equal to?  (it's
\lstinline`xs`).  This information is captured by the following definition by
co\-patterns:%
\footnote{The \lstinline`_,,_` operation is roughly dual to \lstinline`case`:
  whereas \lstinline`case` shallowly inspects a natural number, \lstinline`_,,_`
  shallowly builds a stream.  This is why we refer to this operation as
  \lstinline`cocase` later in \cref{sec:scheme-corec}.}
\begin{contiguous}
\begin{lstlisting}[language=Agda]
  _,,_ : {A : Set} → A → Stream A → Stream A
  head (x ,, xs) = x
  tail (x ,, xs) = xs
\end{lstlisting}
\end{contiguous}
We can write co\-recursive definitions using co\-patterns, too.  For example,
the stream that is \lstinline`always` the same value \lstinline`x`
(\ie \lstinline`x ,, x ,, x ,, ...`)
and the stream built by \lstinline`repeat`edly applying some function
\lstinline`f` to a starting value \lstinline`x`
(\ie \lstinline`x ,, f x ,, f (f x) ,, ...`)
are:
\begin{contiguous}
\begin{lstlisting}[language=Agda]
  always : {A : Set} → A → Stream A
  head (always x) = x
  tail (always x) = always x

  repeat : {A : Set} → (A → A) → A → Stream A
  head (repeat f x) = x
  tail (repeat f x) = repeat f (f x)
\end{lstlisting}
\end{contiguous}
These basic building blocks already let us generate some different streams.  For
example, consider the stream of all zeroes
(\lstinline`0 ,, 0 ,, 0 ,, ...`),
of all the nats
(\lstinline`0 ,, 1 ,, 2 ,, ...`),
and the alternation back and forth between true and false
(\lstinline`true ,, false ,, true ,, false ,, ...`).
Each is an instance of \lstinline|always| and \lstinline|repeat|:
\begin{contiguous}
\begin{lstlisting}[language=Agda]
  zeroes : Stream Nat
  zeroes = always zero
\end{lstlisting}
\end{contiguous}
\begin{contiguous}
\begin{lstlisting}[language=Agda]
  nats : Stream Nat
  nats = repeat succ zero
\end{lstlisting}
\end{contiguous}
\begin{contiguous}
\begin{lstlisting}[language=Agda]
  alt : Stream Bool
  alt = repeat not true
\end{lstlisting}
\end{contiguous}

But not every infinite stream fits within these simple patterns.  For example,
there is the stream that \lstinline`maps` a function \lstinline`f` over every
element of another stream \lstinline`xs`, as described by:
\begin{contiguous}
\begin{lstlisting}[language=Agda]
  maps : {A B : Set} → (A → B) → Stream A → Stream B
  head (maps f xs) = f (head xs)
  tail (maps f xs) = maps f (tail xs)
\end{lstlisting}
\end{contiguous}
\lstinline`maps` is clearly different from \lstinline`always` and
\lstinline`repeat`.  Instead, we need a more general way of generating streams.
We the dual of \lstinline`iter`ation, which is also called an anamorphism
\cite{MFP91}, is defined by co\-patterns as:
\begin{contiguous}
\begin{lstlisting}[language=Agda]
  coiter : {A B : Set} → (B → A) → (B → B) → B → Stream A
  head (coiter f g x) = f x
  tail (coiter f g x) = coiter f g (g x)
\end{lstlisting}
\end{contiguous}

Like \lstinline`repeat g x`, the \lstinline`tail` of the stream is obtained by
updating the seed \lstinline`x` to \lstinline`(g x)`.  Unlike
\lstinline`repeat g x`, each element is calculated on the fly by applying
\lstinline`f` to the seed \lstinline`x`.  Because of this additional step, the
externally observable elements of the stream can be completely different from
the internally private seed used to generate the stream.  Intuitively, the type
\lstinline`B` of the seed is usually ``larger'' than the type \lstinline`A` of
the elements, because each element may be just a small part of the internal
state of the stream.  For example, \lstinline`maps` can be expressed as an
instance of \lstinline`coiter`, where the internal seed is the entire stream of
\lstinline`xs` that have yet to be transformed, each element is obtained by
applying \lstinline`f` to just the \lstinline`head` of the seed, and the seed is
updated with its \lstinline`tail` at every step:
\begin{contiguous}
\begin{lstlisting}[language=Agda]
  maps' : {A B : Set} → (A → B) → Stream A → Stream B
  maps' f xs = coiter (λ xs → f (head xs)) (tail xs)
\end{lstlisting}
\end{contiguous}

Similar to \lstinline`maps` is \lstinline`zipsWith`, which combines the elements
of two streams (pointwise) using some binary function \lstinline`f`.  The
co\-pattern-based definition of \lstinline`zipsWith` is:
\begin{contiguous}
\begin{lstlisting}[language=Agda, basicstyle=\ttfamily\footnotesize]
zipsWith : {A B C : Set} → (A → B → C) → Stream A → Stream B → Stream C
head (zipsWith f xs ys) = f (head xs) (head ys)
tail (zipsWith f xs ys) = zipsWith f (tail xs) (tail ys)
\end{lstlisting}
\end{contiguous}
Is \lstinline`zipsWith` also an instance of \lstinline`coiter`?  Yes, but the
encoding is a little more complex.  Notice that in the \lstinline`tail` case,
the co\-recursive call to \lstinline`zipsWith` is given the same function
\lstinline`f`, but \emph{both} the two stream arguments \lstinline`xs` and
\lstinline`ys` are changed to \lstinline`(tail xs)` and \lstinline`(tail ys)`.
To capture this behavior, the coiteration seed has to contain both streams,
which are modified after every step.  This can be done by pairing the two
together in a product type \lstinline`Stream A × Stream B` as in the following
definition:
\begin{contiguous}
\begin{lstlisting}[language=Agda]
record _×_ A B : Set where
  constructor _,_
  field proj₁ : A
        proj₂ : B
\end{lstlisting}
\end{contiguous}
\begin{contiguous}
\begin{lstlisting}[language=Agda, basicstyle=\ttfamily\footnotesize]
zipsWith' : {A B C : Set} → (A → B → C) → Stream A → Stream B → Stream C
zipsWith' {A} {B} {C} f xs ys = coiter {C} {Stream A × Stream B}
  (λ{(xs , ys) → f (head xs) (head ys)})
  (λ{(xs , ys) → tail xs , tail ys})
  (xs , ys)
\end{lstlisting}
\end{contiguous}
Notice that, while \lstinline`zipsWith'` returns a stream whose elements have
type \lstinline`C`, the internal state used to generate those elements has the
type \lstinline`Stream A × Stream B`, as noted by the explicitly-given implicit
arguments (surrounded in braces \lstinline`{ }`) to \lstinline`coiter`.

Now consider the stream that counts down from a given number to zero, and then
stays at zero from then on.  For example,
\lstinline`countDown 3 = 3 ,, 2 ,, 1 ,, 0 ,, 0 ,, ...`.
This stream can be defined like so, with three main cases to consider:
\begin{contiguous}
\begin{lstlisting}[language=Agda]
  countDown : Nat → Stream Nat
  head (countDown  n      ) = n
  tail (countDown  zero   ) = zeroes
  tail (countDown (succ n)) = countDown n
\end{lstlisting}
\end{contiguous}
The \lstinline`head` of \lstinline`(countdown n)` is always the current number
\lstinline`n`, but the \lstinline`tail` depends on the value of \lstinline`n`.
If it's non-zero, as in \lstinline`tail (countDown (succ n))`, then the stream
proceeds to count down from the predecessor \lstinline`n`.  Otherwise it's zero,
and in this case the rest of the stream is \lstinline`always zero`.  Note that
in this zero case, there is no need to continue counting, because the remaining
elements are already known in advance: the rest of the stream is exactly the
same as \lstinline`zeroes` anyway, so \lstinline`zeroes` itself is returned
instead of referring to \lstinline`countDown` again as in the successor case.
This is more efficient, by avoiding some unnecessary checks against the argument
of \lstinline`countDown`.  Is \lstinline`countDown` an instance of
\lstinline`coiter`?  Yes, this stream contains the same elements:
\begin{contiguous}
\begin{lstlisting}[language=Agda]
  countDown' : Nat → Stream Nat
  countDown' = coiter
    (λ n → n)
    (λ{  zero    → zero
      ; (succ n) → n})
\end{lstlisting}
\end{contiguous}
But this equivalence is \emph{extensional} (only considering what external
observers see), not \emph{intensional} (counting other factors used internally
in the implementation).  In particular, \lstinline`countDown'` will keep
checking the value of the internal seed at every step; even after the seed
becomes 0 and stops changing further, \lstinline`countDown'` will check it at
every \lstinline`tail` projection to find it still the same.

To more accurately reflect the cost of \lstinline`countDown`, we need a more
general way of generating streams.  This is where the \lstinline`corec`ursor
(also known as an apomorphism \cite{Vene98functionalprogramming}) comes into
play: it provides a path for ending co\-recursion once the rest of the stream
becomes fully known in advance.
% Agda is a purely language with no side effects, so it cannot directly model
% both continuations of the co\-inductive step.  However, we can simulate the
We can allow for an early end to co\-recursion by having the co\-inductive type
return a sum (\aka disjoint union \lstinline`B ⊎ Stream A`) of the two possible
options: either return a new seed (of type \lstinline`B`) to continue
co\-recursion, or return a previously-defined stream (of type \lstinline`Stream A`)
to serve as the remainder.
\begin{contiguous}
\begin{lstlisting}[language=Agda]
data _⊎_ A B : Set where
  inj₁ : A → A ⊎ B
  inj₂ : B → A ⊎ B
\end{lstlisting}
\end{contiguous}
\begin{contiguous}
\begin{lstlisting}[language=Agda]
corec : {A B : Set} → (B → A) → (B → Stream A ⊎ B) → B → Stream A
head (corec f g x) = f x
tail (corec f g x) with g x
tail (corec f g x) | inj₁ ys = ys
tail (corec f g x) | inj₂ x' = corec f g x'
\end{lstlisting}
\end{contiguous}
Note that the \lstinline`with` allows us to pattern match on the result of the
update \lstinline`g x`: if it is a stream \lstinline`ys` (the \lstinline`left`
option) then \lstinline`ys` will be returned and co\-recursion stops, but if it
is a new seed \lstinline`x'` (the \lstinline`right` option), then co\-recursion
continues with \lstinline`x'`.  \lstinline`corec` lets us more accurately
capture the intensional details of \lstinline`countDown`, which stops once 0 is
reached.  This definition of \lstinline`countDown` in terms of \lstinline`corec`
is:
\begin{contiguous}
\begin{lstlisting}[language=Agda]
  countDown'' : Nat → Stream Nat
  countDown'' = corec
    (λ n → n)
    (λ{  zero    → inj₁ zeroes
      ; (succ n) → inj₂ n})
\end{lstlisting}
\end{contiguous}

Another example that shows the need to stop the corecursion is the function that
appends a finite \lstinline`List` in front of an infinite stream:
\begin{contiguous}
\begin{lstlisting}[language=Agda]
  data List A : Set where
    []  : List A
    _∷_ : A → List A → List A
\end{lstlisting}
\end{contiguous}
\begin{lstlisting}[language=Agda]
  append : {A : Set} → List A → Stream A → Stream A
  head (append []       ys) = head ys
  head (append (x ∷ xs) ys) = x
  tail (append []       ys) = tail ys
  tail (append (x ∷ xs) ys) = append xs ys
\end{lstlisting}
Note that \lstinline`append` stops co\-recursing when the prefix list is empty.
This behavior---including the end of \lstinline`append`'s co\-recursion---is
correctly expressed by this invocation of \lstinline`corec`:
\begin{lstlisting}[language=Agda]
  append' : {A : Set} → List A → Stream A → Stream A
  append' xs ys = corec
    (λ{ []       → head ys
      ; (x ∷ xs) → x})
    (λ{ []       → left (tail ys)
      ; (x ∷ xs) → right xs})
    xs
\end{lstlisting}

% \begin{lstlisting}[language=Agda]
%   appendForever : {A : Set} → List A → Stream A → Stream A
%   head (appendForever (x :: xs) ys) = x
%   head (appendForever []        ys) = head ys
%   tail (appendForever (x :: xs) ys) = appendForever xs ys
%   tail (appendForever []        ys) = appendForever [] (tail ys)

%   appendForever' : {A : Set} → List A → Stream A → Stream A
%   appendForever' xs ys = coiter
%     (λ{ ((x :: xs) , ys) → x
%       ; ([]        , ys) → head ys})
%     (λ{ ((x :: xs) , ys) → (xs , ys)
%       ; ([]        , ys) → ([] , tail ys)})
%     (xs , ys)
% \end{lstlisting}

%%% Local Variables:
%%% mode: latex
%%% TeX-master: "corec-prog"
%%% End:

\section{Scheme: Resumable Corecursive Control}
\label{sec:scheme-corec}

\lstset{language=Scheme}

Unlike Agda, Scheme does not have built-in support for co\-inductive types or
co\-patterns.  However, we can still represent co\-inductive types as a form of
first-class function which returns a different result depending on the question
it receives (following the technique from \cite{OCamlCopatterns}).%
\footnote{Note that a function is another example of a codata type \cite{CodataInAction}.}
For example, streams can be modeled as a function \lstinline`s` whose single
argument can be one of two options:
\begin{itemize}
\item If \lstinline`s` is passed \lstinline`'head`, then the first element is
  returned.
\item If \lstinline`s` is passed \lstinline`'tail`, then the rest of the stream
  is returned.
\end{itemize}
Informally, the type of these functions can be described by the following
specification, which is an \emph{intersection} (written as \lstinline`&`)
between two more specific function types:
\begin{contiguous}
\begin{lstlisting}[language=Scheme]
  ;; Stream a = 'head -> a
  ;;          & 'tail -> Stream a
\end{lstlisting}
\end{contiguous}
The function type \lstinline`'head -> a` denotes functions which can only be
applied to exactly the symbol \lstinline`'head`, and returns some \lstinline`a`
result.  Likewise, a function \lstinline`'tail -> Stream a` can only be applied
to \lstinline`'tail`.  In total, both of these requirements together say that a
\lstinline`Stream a` is any function that returns a value of type \lstinline`a`
when applied to \lstinline`'head`, and returns another \lstinline`Stream a` when
applied to \lstinline`'tail`.

The \lstinline`Stream a` interface lets us define a number of functions that
operate over any generic \lstinline`Stream`.  For example, an operations that
\lstinline`takes` a number of elements from a stream, \lstinline`drops` a number
of elements from the start of a stream, and fetches the element at an
\lstinline`index` can each be defined like so:
\begin{contiguous}
\begin{lstlisting}[language=Scheme]
  ;; takes : (Stream a, Nat) -> List a
  (define (takes s n)
    (cond [(= n 0) '()]
          [(= n 1) (list (s 'head))]
          [else (cons (s 'head) (takes (s 'tail) (- n 1)))]))
\end{lstlisting}
\end{contiguous}
\begin{contiguous}
\begin{lstlisting}[language=Scheme]
  ;; drops : (Stream a, Nat) -> Stream a
  (define (drops s n)
    (cond [(= n 0) s]
          [else (drops (s 'tail) (- n 1))]))
\end{lstlisting}
\end{contiguous}
\begin{contiguous}
\begin{lstlisting}[language=Scheme]
  ;; index : (Stream a, Nat) -> a
  (define (index s n) ((drops s n) 'head))
\end{lstlisting}
\end{contiguous}
But each of these functions assumes we have a stream already.  How can we create
a new stream from scratch in Scheme?  Streams can be created by just returning 
any first class function matching the \lstinline`Stream` interface.  For
example, the infinite stream of zeros, fitting this type specification, is:
\begin{contiguous}
\begin{lstlisting}[language=Scheme]
  (define zeroes
    (lambda (question)
      (cond [(equal? question 'head) 0]
            [(equal? question 'tail) zeroes])))
\end{lstlisting}
\end{contiguous}
So that no matter how deep you go,
\lstinline`(((((zeroes 'tail) 'tail) 'tail) ... ) 'head)`
will always return \lstinline`0`.

The definition of \lstinline`zeroes` uses a structure (a \lstinline`lambda`
taking the caller's question, then testing if that question is equal to
\lstinline`'head` or \lstinline`'tail`) that is extremely common for creating
streams.  Thankfully, we can write a macro which abstracts over this structure
to make streams easier to write in Scheme.  The \lstinline`cocase` macro
introduces syntactic sugar for creating a stream by cases on the observation:%
\footnote{\lstset{basicstyle=\ttfamily\footnotesize}%
  Using \lstinline`(cocase ['head x] ['tail xs])` to generate a stream is
  equivalent to \lstinline[language=Agda]`x ,, xs` from \cref{sec:agda-corec}.}
\begin{contiguous}
\begin{lstlisting}[language=Scheme]
  (define-syntax-rule (cocase [destructor result] ...)
    (lambda (question)
      (cond [(equal? question destructor) result]
            ...)))
\end{lstlisting}
\end{contiguous}
So that this alternative definition of \lstinline`zeroes` expands exactly into
the longer one above.
\begin{lstlisting}[language=Scheme]
  (define zeroes (cocase ['head 0] ['tail zeroes]))
\end{lstlisting}
We can also
generalize \lstinline`zeroes` to any stream that \lstinline`always` returns the
same value \lstinline`x`:
%\begin{lstlisting}[language=Scheme]
%  (define (always x)
%    (letrec [(xs (cocase ['head x] ['tail xs]))]
%      xs))
%\end{lstlisting}

\begin{lstlisting}[language=Scheme]
  (define (always x) (cocase ['head x] ['tail (always x)]))
\end{lstlisting}

As with \lstinline`zeroes`, we can use \lstinline`cocase` and self-reference to
create a wide variety of infinite streams.  For example, the streams which count
up starting from some given number \lstinline`n`
(\lstinline`n (+ n 1) (+ n 2) ...`),
or down
(\lstinline`n (- n 1) ...  1 0 0 ...`; staying at 0)
are:
\begin{contiguous}
\begin{lstlisting}[language=Scheme]
  (define (count-up n)
    (cocase ['head n]
            ['tail (countUp (+ n 1))]))
\end{lstlisting}
\end{contiguous}
\begin{contiguous}
\begin{lstlisting}[language=Scheme]
  (define (count-down n)
    (cocase ['head n]
            ['tail (cond [(= n 0) zeroes]
                         [else (countDown (- n 1))])]))
\end{lstlisting}
\end{contiguous}
Notice how the three clauses defining the analogous \lstinline`countDown` via
co\-patterns in Agda (\cref{sec:agda-corec}) can be found in the definition
\lstinline`count-down` here.

Just like before, we can abstract out a common pattern of generating streams by
\lstinline`coiter`ation, which uses an internal \lstinline`state` to
\lstinline`make` the \lstinline`'head` on the fly.  To generate the
\lstinline`'tail`, we need to \lstinline`update` the \lstinline`state` and
continue \lstinline`coiter`ating.
\begin{contiguous}
\begin{lstlisting}[language=Scheme]
  ;; coiter : (b -> a, b -> b, b) -> Stream a
  (define (coiter make update state)
    (cocase
     ['head (make state)]
     ['tail (coiter make update (update state))]))
\end{lstlisting}
\end{contiguous}
As we know, the operation which \lstinline`maps` a function \lstinline`f` over
all the elements of a stream is an instance of \lstinline`coiter`ation in Scheme, too.
\begin{contiguous}
\begin{lstlisting}[language=Scheme]
  ;; maps* : (a -> b, Stream a) -> Stream b
  (define (maps* f xs)
    (coiter
     (lambda (xs) (f (xs 'head)))
     (lambda (xs) (xs 'tail))
     xs))
\end{lstlisting}
\end{contiguous}

But not every stream can be generated from just \lstinline`coiter`ation.  We
have seen in the previous section that even simple functions like
\lstinline`count-down` are not faithfully captured by coiteration.  In Agda, we
used a sum type that lets the programmer control when co\-recursion ends early
(possibly continuing forever).  But once a \lstinline`corec` loop is ended in
this way, it is done for good.  In Scheme, we can do something more: using the
first-class control provided by \lstinline`call/cc`, we can provide two
continuations in the \lstinline`'tail` step of the co\-recursive loop.  The
first (implicit) continuation lets the \lstinline`'tail` step update the state
of the loop and continue co\-recursing, while the second (explicit) continuation
captures the caller who requested the \lstinline`'tail` of the stream.  As such,
we define the \emph{classical} \lstinline`corec` as the following generalization
of \lstinline`coiter`:
\begin{contiguous}
\begin{lstlisting}[language=Scheme]
  ;; corec : (b -> a, (Cont (Stream a), b) -> b, b) -> Stream a
  (define (corec make update state)
    (cocase
     ['head (make state)]
     ['tail (call/cc
              (lambda (finish)
                (corec make update (update finish state))))]))
\end{lstlisting}
\end{contiguous}
Unlike a sum type which definitively ends the loop once and for all,
continuations can be invoked multiple times, which give the ability to ``pause''
and ``resume'' the loop at will.  Yet, the only difference from
\lstinline`coiter` above is that in the co\-inductive step, \lstinline`corec`
captures the \lstinline`'tail` caller in the continuation \lstinline`finish`,
and provides it to the \lstinline`update` function along with the current
\lstinline`state`.  Other than this difference, \lstinline`corec` and
\lstinline`coiter` are the same.

Analogous to appending a list in front of a stream in Agda
(\cref{sec:agda-corec}), the direct definition of \lstinline`append` in Scheme
is:
\begin{contiguous}
\begin{lstlisting}[language=Scheme]
  ;; append : (List a, Stream a) -> Stream a
  (define (append xs ys)
    (cocase ['head (cond [(null? xs) (ys 'head)]
                         [else (car xs)])]
            ['tail (cond [(null? xs) (ys 'tail)]
                         [else (append (cdr xs) ys)])]))
\end{lstlisting}
\end{contiguous}
This \lstinline`append` function can alternatively be  defined as an instance of
\lstinline`corec`:
\begin{contiguous}
\begin{lstlisting}[language=Scheme]
  (define (append* xs ys)
    (corec xs
     (lambda (xs)
       (cond [(null? xs) (ys 'head)]
             [else (car xs)]))
     (lambda (finish xs)
       (cond [(null? xs) (finish (ys 'tail))]
             [else (cdr xs)]))))
\end{lstlisting}
\end{contiguous}
The state of this \lstinline`corec`ursion is the finite prefix list
\lstinline`xs`.  Notice that the bodies of the two functions in
\lstinline`append*` are almost identical to the \lstinline`'head` and
\lstinline`'tail` branches used in \lstinline`append`.  The difference is that
the co\-inductive step, instead of returning directly when \lstinline`xs` is
empty, must invoke the continuation \lstinline`finish` to end the loop.  And
instead of continuing the loop by calling itself when \lstinline`xs` is
non-empty, we simply return the updated state \lstinline`(cdr xs)`.

%\subsubsection*{Expressiveness of control and corecursion}
\subsection{The power of classical corecursion}
\label{sec:classical-expressiveness}

The use of classical \lstinline`corec` to encode \lstinline`append` 
amounts to the same as the non-classical version in Agda (\cref{sec:agda-corec}).
There is an improved performance from ending the loop early, but performance
aside, the same result could be calculated via \lstinline`coiter` instead.

However, sometimes the difference between \lstinline`coiter`ation and
\lstinline`corec`ursion goes beyond just mere performance.  Some streams truly
need the full generality of a classical \lstinline`corec` to be defined at
all. In here, we exhibit the expressive power of classical corecursion over
coiteration in a language like Scheme with control operators.%
\footnote{In a purely functional language, classical \lstinline`corec` is
  clearly more expressive than \lstinline`coiter`: it captures and provides a
  first-class continuation to the \lstinline`'tail` branch, which is otherwise
  not possible in a pure language.}
In particular, the access to the continuation pointing to the \lstinline`'tail`
caller gives more flexibility than merely stopping the iteration.  For example,
consider this fact about infinite bit streams made up of just two different
element values \lstinline`#t` (the 1 bit) and \lstinline`#f` (the 0 bit):
\begin{fact}
\label{thm:infinite-stream-pigeons}

For every infinite stream $s$ over a finite alphabet $A$, there is an element
$x \in A$ such that $x$ occurs infinitely often in $s$.
As a consequence, every infinite bit stream (over the alphabet \lstinline`#t`
and \lstinline`#f`) contains either an infinite number of \lstinline`#t`s
or an infinite number of \lstinline`#f`s.
\end{fact}
This is an application of the pigeon-hole property.  If there are only two holes
(\lstinline`#t` or \lstinline`#f`) that together have to house an infinite
number of occupants, than at least one of the holes (maybe both, but not
necessarily) must have an infinite number of occupants.  And this fact
generalizes to any finite alphabet, because that still leads to a finite number
of holes being filled with an infinite number of occupants.

Given a bit stream, can we calculate outright which value occurs infinitely
often?  Unfortunately, no.%
\footnote{We could be given the stream that is \lstinline`always #f`, so the
  answer is obviously \lstinline`#f`, but we could be given the stream of 100
  \lstinline`#f`s before it is \lstinline`always #t`, and we would need to know
  the answer is \lstinline`#t` even though there are many \lstinline`#f`s at the
  start.
  % We could also be given the stream of 100 \lstinline`#f`s, followed by 1
  % million \lstinline`#t`s, and \emph{then} is \lstinline`always #f`.
  There is no way to know, \emph{a priori}, how deep into a stream we need to
  check to be confident which element appears infinitely often, and we cannot
  exhaustively check all of them.}
However, we can weaken \cref{thm:infinite-stream-pigeons} slightly: instead of
an infinite number of occurrences of $x$ given all at once, we can promise only
finite occurrences but in any amount desired.%
\begin{fact}
\label{thm:approx-stream-pigeons}

For every infinite stream $s$ over a finite alphabet $A$ (such as
$A = \{\text{\tt\lstinline`#t`}, \text{\tt\lstinline`#f`}\}$), and for any $n$,
there is an element $x \in A$ such that there are $n$ different occurrences of
$x$ in $s$.
\end{fact}
%\zena{Should we say n>0?} \paul{It's true for $n=0$, too.}
This fact is much more tractable.  Up front, we are only asked for a specific
amount $n$ of occurrences, so we know when we have gathered enough evidence to
say which element occurs that many times.  Concretely, we can represent this
evidence as indexes $i_0, i_1, \dots $ into the original stream $s$, such that
the value of $s$ at each such $i$ is the same $x$.  Now, the main challenge is
that the observer can ask for many different number of occurrences, and so our
options for which $x$ is chosen might have to change as that count increases.
For example, consider the stream $s$ made up of 100 \lstinline`#f`s, followed
by 1 million \lstinline`#t`s, and then infinite \lstinline`#f`s.  If we are
asked for an element that appears 10 times, then we might say \lstinline`#f`,
because it is very common at the start of the list.  But then if we are asked
for 1000 occurrences, we might want to say \lstinline`#t` because many more of
them will be found before we see the $101^{st}$ \lstinline`#f`.  Yet, if we are
asked for 1 billion occurrences, we have no choice but to say \lstinline`#f`;
there simply aren't enough \lstinline`#t` occurrences in $s$ to satisfy the
request.

We can mediate between \cref{thm:infinite-stream-pigeons} and
\cref{thm:approx-stream-pigeons} using first-class control in Scheme.
Effectively, we can provide a stream that appears to implement
\cref{thm:infinite-stream-pigeons} to the programmer.  Yet, at the end of the
day, only \cref{thm:approx-stream-pigeons} need be implemented, because every
terminating program can only inspect a finite number of elements in a stream
before it ends.  The first-class control present in classical co\-recursion lets
us automatically infer this finite number while the program runs, without the
programmer's explicit knowledge or intervention.

The \lstinline`call/cc` operator creates a check-point by capturing our observer
in a continuation that we can invoke several times, back-tracking to the point
in time when \lstinline`call/cc` was called so we can provide several different
answers to that same observer.  In this application, we can start to look for
the infinite common occurrences by first creating a check-point with
\lstinline`call/cc`, and then \emph{guessing} that the head element of the
stream is the bit that might occur infinitely often.  As long as we keep finding
more repetitions of that head bit, then our guess appears correct, and we can
keep providing more indexes to repetitions of that bit.  However, if we find the
other bit in the stream, then our guess might be \emph{wrong}.  In this case, we
can back-track to the start and change our answer to the other bit, continuing
to search into the remainder of the stream.  If we find a repetition of the
first bit again, we can back-track yet again to where we left off originally,
rather than starting over entirely.

\begin{figure}
\centering
\begin{lstlisting}[language=Scheme,basicstyle=\ttfamily\footnotesize]
;; infinite-bits : Stream a -> Stream Nat
;;     where `a` is a 2-value type (like Bool = #t | #f)
(define (infinite-bits s)
  (call/cc
   (lambda (restart)
     (infinite-bit0 (s 'head) (s 'tail) 0 restart))))
\end{lstlisting}
\begin{lstlisting}[language=Scheme,basicstyle=\ttfamily\footnotesize]
;; infinite-bit0 : (a, Stream a, Nat, Cont (Stream a)) -> Stream Nat
(define (infinite-bit0 bit0 rest depth switch)
  (cocase
   ['head depth]
   ['tail
    (cond
      [(equal? (rest 'head) bit0)
       (infinite-bit0 bit0 (rest 'tail) (+ 1 depth) switch)]
      [else
       (call/cc
        (lambda (return)
          (switch
            (infinite-bit1 bit0 (rest 'tail) (+ 1 depth) return))))])]))
\end{lstlisting}
\begin{lstlisting}[language=Scheme,basicstyle=\ttfamily\footnotesize]
;; infinite-bit1 : (a, Stream a, Nat, Cont (Stream a)) -> Stream Nat
(define (infinite-bit1 bit0 rest depth switch)
  (cocase
   ['head depth]
   ['tail
    (cond
      [(not (equal? (rest 'head) bit0))
       (infinite-bit1 bit0 (rest 'tail) (+ 1 depth) switch)]
      [else
       (call/cc
        (lambda (return)
          (switch
            (infinite-bit0 bit0 (rest 'tail) (+ 1 depth) return))))])]))
\end{lstlisting}
\caption{Search for infinite repetitions of a bit in a bit stream.}
\label{fig:infinite-bits}
\end{figure}

This algorithm can be implemented in Scheme as shown in
\cref{fig:infinite-bits}.  The top-level \lstinline`infinite-bits` function
takes any stream \lstinline`s` made up of only two different elements (like
\lstinline`#t` and \lstinline`#f` or \lstinline`'a` and \lstinline`'b`).  For
the sake of distinguishing these two options, the first element encountered at
the \lstinline`'head` of \lstinline`s` is considered the 0 bit; the other one
not seen yet is the 1 bit.  The task of \lstinline`infinite-bits` is to return a
stream of natural number indices into \lstinline`s` all pointing to the same
bit.  To begin, \lstinline`infinite-bits` invokes \lstinline`call/cc` to save a
check-point for when it was first called, making it possible to completely
\lstinline`restart` the stream over from the very beginning, if needed.  Then,
\lstinline`infinite-bits` guesses that there will be enough repetitions of the 0
bit and attempts to find them using the first helper function
\lstinline`infinite-bit0`:
\begin{itemize}
\item The first argument \lstinline`(s 'head)` is the value of the 0 bit.
\item The second argument \lstinline`(s 'tail)` is the stream we are searching.
\item The third argument keeps track of the depth we have descended into
  \lstinline`s`, used for calculating the indexes.  Since we begin at the start
  of the stream, the initial value is \lstinline`0`.
\item The last argument \lstinline`restart` is a continuation that lets us
  replace our initial guess with the other possibility: that there are infinite
  repetitions of the 1 bit in the stream.
\end{itemize}
The \lstinline`'head` of \lstinline`infinite-bit0` is the current
\lstinline`depth` that we have searched into the original \lstinline`s`.  The
\lstinline`'tail` is computed on demand, and it depends on the \lstinline`'head`
element of the \lstinline`rest` of \lstinline`s`:
\begin{itemize}
\item If it the 0 bit we're looking for, then we just continue
  looking for more occurrences of it in the \lstinline`'tail` of
  \lstinline`rest`, making sure to increment our \lstinline`depth`.
\item Otherwise, it is the 1 bit.  In this case, we pause our current search for
  0 bits, change our guess to say that the 1 bit occurs infinitely often
  instead, and begin looking for 1s by switching to the other helper function
  \lstinline`infinite-bit1`.  In order to perform the switch, we have to create
  another check-point saving our current progress in the search for 0 bits.
  This check-point is another continuation captured by \lstinline`call/cc`,
  which is passed to \lstinline`infinite-bit1` so that it might switch back and
  resume the search for 0 bits from where it was paused.
\end{itemize}
The second helper function \lstinline`infinite-bit1` is defined almost
identically to \lstinline`infinite-bit0`.  The only difference is that when
checking each \lstinline`'head` element of the \lstinline`rest` of the stream,
\lstinline`infinite-bit1` instead checks for elements that are \emph{not} equal
to the 0 bit; by the process of elimination, these must be the 1 bit.

The main property of \lstinline`infinite-bits` follows the logic of
\cref{thm:infinite-stream-pigeons}: For any stream \lstinline`s`, the lookup
\lstinline`(index s i)` always returns the same bit \lstinline`x` for every
index \lstinline`i` in \lstinline`(infinite-bits s)`.
\begin{property}[Infinite Bits --- Binary]
\label{prop:infinite-bits-binary}

Under the pre-condition that \lstinline`infinite-bits` is given an infinite bit
stream \lstinline`s`, its post-condition is that there is a bit value
\lstinline`x` such that
\begin{lstlisting}
  (maps (lambda (i) (index s i)) (infinite-bits s))
\end{lstlisting}
is observationally equivalent to \lstinline`(always x)`.  
\end{property}
For example, we can test out \lstinline`infinite-bits` by \lstinline`append`ing
some irregular variations on top of a stream that is \lstinline`always` some
constant value.  If we ask for only 3 repeated occurrences in the stream
\lstinline`#t #f #f #t #f #t #t #t ...`
then \lstinline`infinite-bits` will first find 3 occurrences of \lstinline`#f`
before anything else, pointing out their indexes at 1, 2, and 4:
\begin{contiguous}
\begin{lstlisting}[language=Scheme]
(takes (infinite-bits (append '(#t #f #f #t #f) (always #t))) 3)
= '(1 2 4)
\end{lstlisting}
\end{contiguous}
However, if we ask for 5 repetitions in that very same stream, there are simply
not enough \lstinline`#f`s to be found.  Thus, \lstinline`infinite-repetitions`
will point out 5 different indexes to \lstinline`#t`s:
\begin{contiguous}
\begin{lstlisting}[language=Scheme]
(takes (infinite-bits (append '(#t #f #f #t #f) (always #t))) 5)
= '(0 3 5 6 7)
\end{lstlisting}
\end{contiguous}
Note that, despite the fact that the answer might depend on the observation, a
single call to \lstinline`infinite-bits` will always give \emph{consistent}
answers throughout its lifetime no matter how many times the result is
inspected.  For example, if we apply both of the above tests to the
\emph{same} stream returned by \lstinline`infinite-bits`, we get consistent
approximations each time.
\begin{contiguous}
\begin{lstlisting}[language=Scheme]
(let [(ix (infinite-bits (append '(#t #f #f #t #f) (always #t))))]
  (list (takes ix 3) (takes ix 5)))
= '((0 3 5) (0 3 5 6 7))
\end{lstlisting}
\end{contiguous}
Even though the first test (asking for only 3 indexes) could initially produce
the different approximate result \lstinline`'(1 2 4)` shown above, it is
automatically updated with \lstinline`'(0 3 5)` to be consistent with the
second, more strenuous, test (asking for 5 indexes).

\begin{figure}
\centering
\begin{lstlisting}[language=Scheme,basicstyle=\ttfamily\footnotesize]
  ;; infinite-bits* : Stream a -> Stream Nat
  (define (infinite-bits* s)
    (call/cc
     (lambda (restart)
       (infinite-bit0* (s 'head) (s 'tail) 0 restart))))
\end{lstlisting}
\begin{lstlisting}[language=Scheme,basicstyle=\ttfamily\footnotesize]
  ;; infinite-bit0* : (a, Stream a, Cont (Stream Nat)) -> Stream Nat
  (define (infinite-bit0* bit0 rest depth switch)
    (coiter
     (list rest depth switch)
     (lambda (state) (match state [(list _ depth _) depth]))
     (lambda (state)
       (match state
         [(list rest depth switch)
          (cond
            [(equal? bit0 (rest 'head))
             (list (rest 'tail) (+ 1 depth) switch)]
            [else
             (call/cc
              (lambda (resume)
                (switch
                 (infinite-bit1*
                  bit0 (rest 'tail) (+ 1 depth) resume))))])]))))
\end{lstlisting}
\begin{lstlisting}[language=Scheme,basicstyle=\ttfamily\footnotesize]
  ;; infinite-bit1* : (a, Stream a, Nat,
  ;;                    Cont (Stream a * Nat * Cont (Stream Nat)))
  ;;                -> Stream Nat
  (define (infinite-bit1* bit0 rest depth switch)
    (corec
     (list rest depth switch)
     (lambda (state) (match state [(list _ depth _) depth]))
     (lambda (return state)
       (match state
         [(list rest depth switch)
          (cond
            [(not (equal? bit0 (rest 'head)))
             (list (rest 'tail) (+ 1 depth) switch)]
            [else (switch (list (rest 'tail) (+ 1 depth) return))])]))))
\end{lstlisting}
\caption{Searching for infinite repetitions in a bit-stream using \lstinline`corec`.}
\label{fig:infinite-bits-corec}
\end{figure}

As we alluded to, identifying infinite repetitions cannot be an instance of
\lstinline`coiter`.  That's because the crucial \lstinline`infinite-bit0` and
\lstinline`infinite-bit1` helper functions need to capture the caller of each
\lstinline`'tail` request (with \lstinline`call/cc`), in order to save new
check-points during the search.  \lstinline`coiter`---even when combined with
\lstinline`call/cc`---cannot generate \lstinline`infinite-bits` because it hides
the \lstinline`'tail` caller from the \lstinline`update` function used in the
\lstinline`'tail` step.  Notice that if we invoke \lstinline`coiter` as
\begin{lstlisting}[language=Scheme]
  (coiter make (lambda (new-state) (call/cc (k) ...)) state)
\end{lstlisting}
then the continuation \lstinline`k` captures the context which updates
\lstinline`coiter`'s state, \emph{not} the \lstinline`'tail` caller of
\lstinline`coiter`.  Instead, \lstinline`corec` provides exactly this extra
information to \lstinline`update`.  Because of this, we can implement
\cref{thm:infinite-stream-pigeons} using two nested \lstinline`corec`ursive
loops, as shown in \cref{fig:infinite-bits-corec}.  The two top-level functions
\lstinline`infinite-bits` and \lstinline`infinite-bits*` are identical.  The
outer loop of \lstinline`infinite-bits*` looks for occurrences of bit 0 (the one
we found first at the \lstinline`'head` of the stream).  The inner loop(s) looks
for occurrences of bit 1 (those not equal to bit 0).

The outer loop \lstinline`infinite-bit0` can be implemented as an instance of
\lstinline`coiter`, where the state is:
\begin{enumerate*}
\item the \lstinline`rest` of the stream to search through,
\item the current \lstinline`depth` into the original stream, and
\item a continuation to \lstinline`switch` to the other searching mode.
\end{enumerate*}
The value of what we are searching for---the bit 0 that appeared at the
\lstinline`'head` of the original stream---is not part of the state, because it
doesn't change.  As before, the \lstinline`'head` element of this stream is the
current depth (found in the state), and the \lstinline`'tail` is computed
on-demand.  If the next element is equal to 0 bit, then we just continue
\lstinline`coiter`ating with an updated state with 1 more depth and with the
\lstinline`'tail` of the \lstinline`'rest` of the stream.  Otherwise, we found a
1 bit (because it is different from the 0 bit), and we have to
\lstinline`switch` our searching mode to look for more 1 bits.  To do so, we
save our place in this outer loop (with a call to \lstinline`call/cc`) so that
we can \lstinline`resume` it later, and invoke the \lstinline`switch` to search
for 1 bits with a new inner loop.

The inner loop \lstinline`infinite-bit1` looks for 1 bits (that are not equal to
the 0 bit), and is an instance of \lstinline`corec`.  The state of this
\lstinline`corec`ursion is similar to that of \lstinline`infinite-bit0` except
that the type of the \lstinline`switch`ing continuation is different: rather
than a fully-formed stream of indexes, it expects a new state that can be used
to update and continue the outer-loop.  So long as \lstinline`infinite-bit1`
finds more 1 bits, it will continue its loop by updating its state in the
co\-inductive step as above.  But once \lstinline`infinite-bit1` finds a 0 bit,
it has to \lstinline`switch` back to the outer-loop, kick-starting it back up.
To do so, it passes an updated state including the new continuation that
\lstinline`return`s to this \lstinline`'tail` caller.

\begin{figure}
\centering
\begin{lstlisting}[language=Scheme,basicstyle=\ttfamily\footnotesize]
;; infinite-repetitions : Stream a -> Stream Nat
(define (infinite-repetitions s)
  (call/cc
   (lambda (start)
     (let [(restart (lambda (y indexes) (start indexes)))]
       (infinite-of (s 'head) (s 'tail) 0 restart)))))
\end{lstlisting}
\begin{lstlisting}[language=Scheme,basicstyle=\ttfamily\footnotesize]
;; infinite-of : (a, Stream a, Nat, Cont (a * Stream Nat)) -> Stream Nat
(define (infinite-of x rest depth switch)
 (cocase
  ['head depth]
  ['tail
   (let [(next (rest 'head))]
    (cond
     [(equal? next x)
      (infinite-of x (rest 'tail) (+ 1 depth) switch)]
     [else
      (call/cc
       (lambda (return)
        (let [(resume
               (lambda (y indexes)
                 (cond [(equal? x y) (return indexes)]
                       [else (switch y indexes)])))]
          (switch next
           (infinite-of next (rest 'tail) (+ 1 depth) resume)))))]))]))
\end{lstlisting}
\caption{Search for infinite repetitions of elements in any stream.}
\label{fig:infinite-repetitions}
\end{figure}

Thus far, we have only considered bit streams made up of only two different
values.  Yet, \cref{thm:infinite-stream-pigeons,thm:approx-stream-pigeons} both
promise to work with any finite alphabet, not just binary ones.  What happens if
we try to apply \lstinline`infinite-bits`---or equivalently
\lstinline`infinite-bits*`---to some other non-bit stream?  This does not
satisfy the pre-condition of the main defining \cref{prop:infinite-bits-binary}
for this algorithm.  Instead, we can only ensure this weakened version:
\begin{property}[Infinite Bits --- $n$-ary]
\label{prop:infinite-bits-n-ary}

Given \emph{any} infinite stream \lstinline`s`,
\begin{lstlisting}
  (maps (lambda (i) (index s i)) (infinite-bits s))
\end{lstlisting}
is observationally equivalent to one of the two following streams:
\begin{enumerate}
\item \lstinline`(always (s 'head))`, or
\item some stream of elements \lstinline`x1 x2 x3 ...` all non-equal to
  \lstinline`(s 'head)`.
\end{enumerate}
\end{property}
So in the general case, \lstinline`infinite-bits` might not deliver what we were
expecting.  Given the stream \lstinline`'a #t #f #t #f ...`,
\lstinline`infinite-bits` returns the indexes \lstinline`1 2 3 4...`
pointing out the alternating sequence of \lstinline`#t` and \lstinline`#f`
values---all different from the first element \lstinline`'a`.

If we want to generalize \lstinline`infinite-bits` to search streams for
infinite repetitions among any number of different elements, we need to move
beyond the binary distinction made between its two helper functions.  This
generalization is made in \cref{fig:infinite-repetitions}.  The main difference
can be seen in the continuation that is used by the single helper function
\lstinline`infinite-of` that searches for repetitions of any given value
\lstinline`x`.  In addition to a new stream of \lstinline`indexes`, this
continuation also expects the value \lstinline`y` that each index points to.
The extra information associates each element value \lstinline`y` with the
current progress in the search for \lstinline`y` occurrences.  In effect, this
builds an association map between the different element values that make up the
stream and the different continuations waiting for the stream of indexes to
those values.  The helper function \lstinline`infinite-of` can then
\lstinline`switch` which element it is searching for by invoking the
continuation with the \lstinline`next` element it found.  In order to return
back to this place in the paused search, \lstinline`call/cc` is used to save the
\lstinline`return` continuation to the current \lstinline`'tail` caller, and the
\lstinline`switch`ing continuation is updated by associating the current value
\lstinline`x` with this \lstinline`return` (with all other values \lstinline`y`
remaining unchanged).

We can almost write \lstinline`infinite-repetitions` in terms of
\lstinline`corec`.  There is just one problem: \lstinline`infinite-repetitions`
might not terminate!  If \lstinline`infinite-repetitions` is given a stream over
an infinite alphabet (like, say, \lstinline`Stream Nat`), then there might not
be any infinite repetitions.  For example, we cannot point to two occurrences of
the same element in \lstinline`(count-up 0)`, and
\lstinline`infinite-repetitions` will loop forever if it tries.  This
illustrates one utility of \lstinline`corec` and \lstinline`coiter`: they cannot
cause infinite loops.  In that way, we know that \lstinline`infinite-bits`
terminates because its manual recursion it can be encoded away as
\lstinline`infinite-bits*` in terms of only \lstinline`coiter` and
\lstinline`corec`.  Though the reason for termination may not be entirely
obvious on the surface, these combinators bake it in syntactically.  And this is
why we can't encode \lstinline`infinite-repetitions` in the same way.  The
pre-condition to \lstinline`infinite-repetitions`---that the stream be built
from a finite alphabet---is implicit, so the termination condition cannot be
easily expressed in terms of \lstinline`corec`.

% \begin{fact}
% \label{thm:first-or-other-pigeons}
%
% For every infinite stream $s$ over any alphabet $A$ (even infinite ones),
% either $head(s)$ occurs infinitely often in $s$, or $s$ contains an infinite
% number of other elements which are different from $head(s)$.
% \end{fact}

%%% Local Variables:
%%% mode: latex
%%% TeX-master: "corec-prog"
%%% End:

\section{Python: Corecursive Objects and Exceptions}
\label{sec:python-corec}

\lstset{language=Python}

\begin{figure}
\centering
\begin{lstlisting}[language=Python]
class Stream:
    # head : Stream -> elem
    # tail : Stream -> Stream

    def __iter__(self):
        curr = self
        while True:
            yield curr.head()
            curr = curr.tail()

    def drops(self, n):
        curr = self
        for i in range(n):
            curr = curr.tail()
        return curr

    def takes(self, n):
        return [ elem for (elem, i) in zip(self, range(n)) ]

\end{lstlisting}
\caption{An abstract class of infinite \lstinline`Stream`s in Python.}
\label{fig:python-infinite-stream}
\end{figure}

So far, we've seen co\-recursion over streams in two different languages---Agda
and Scheme---which are both functional.  Does that mean that co\-recursion only
makes sense in the context of the functional paradigm?  No!  In fact, the idea
of co\-recursion over co\-inductive types maps closely to familiar concepts in
the object-oriented paradigm.  In particular, co\-inductive types correspond to
an interface, and the model of co\-recursion we have seen thus far corresponds
to a form of \emph{immutable} objects.  For example, consider how
streams---objects defined in terms of their \lstinline`head` and
\lstinline`tail` projections---can be defined as the abstract Python class shown
in \cref{fig:python-infinite-stream}.  The actual response of the fundamental
\lstinline`head` and \lstinline`tail` depend on the specific stream in question,
and cannot be defined in the general case.  Thus, they are left abstract for
now.  However, some helpful derived operations can be given for any stream---so
long as they can be defined only in terms of \lstinline`head` and
\lstinline`tail`.  For example, we can give a method that \lstinline`drops` a
given number of elements from the front of the stream, and one that
\lstinline`takes` the first \lstinline`n` elements and returns them in a list.
Other special methods that are expected in Pythonic style---such as the
\lstinline`__iter__`method for iterating sequentially through the elements of
the stream---can also be given a generic definition via \lstinline`head` and
\lstinline`tail`.%
\footnote{The abstract \lstinline`Stream` class can be seen as another instance
  of a codata type \cite{CodataInAction}, where the interface corresponds to a
  codata type definition.  Default methods of the \lstinline`Stream` interface
  in \cref{fig:python-infinite-stream} correspond to operations on objects of
  the codata type.  And the following subclasses of \lstinline`Stream` that
  define the \lstinline`head` and \lstinline`tail` methods are all functions
  which create objects of the \lstinline`Stream` codata type.}

If the abstract class \lstinline`Stream` defines the type of streams, then how
do we define actual stream objects?  These are given through subclasses of
\lstinline`Stream` that give real implementations of the \lstinline`head` and
\lstinline`tail` methods.  For example, the class of streams generated by
\lstinline`Repeat`ing the same function on an initial value is:
\begin{contiguous}
\begin{lstlisting}[language=Python]
class Repeat(Stream):
    def __init__(self, state, update=lambda x: x):
        self.state = state
        self.update = update

    def head(self):
        return self.state
    def tail(self):
        return Repeat(self.update(self.state), self.update)
\end{lstlisting}
\end{contiguous}
The \lstinline`head` of \lstinline`Repeat(x,f)` is just \lstinline`x`.  The
\lstinline`tail` is calculated by applying the \lstinline`update` function to
the \lstinline`state`.  That way \lstinline`Repeat(x,f).tail()` returns
\lstinline`Repeat(f(x),f)`, so \lstinline`Repeat(x,f)` simulates the infinite
stream
\lstinline`x, f(x), f(f(x)), f(f(f(x))), ...`.
By default, the \lstinline`update` function returns its input unchanged, so
\lstinline`Repeat(x)` will just repeat \lstinline`x` forever.  Given:
\begin{contiguous}
\begin{lstlisting}[language=Python]
  zeroes = Repeat(0)
  nats = Repeat(0, lambda x: x+1)
\end{lstlisting}
\end{contiguous}
then \lstinline`zeroes` simulates the stream \lstinline`0, 0, 0, 0 ...` while
\lstinline`nats` simulates \lstinline`0, 1, 2, 3, ...`.

The class of \lstinline`CoIter`ative objects generalizes \lstinline`Repeat` to
somehow \lstinline`make` each element from the current value of the
\lstinline`state`, rather than requiring those elements are just the
\lstinline`state` exactly as-is.  This generalization is given as this alternate
subclass of \lstinline`Stream`:
\begin{contiguous}
\begin{lstlisting}[language=Python,basicstyle=\ttfamily\footnotesize]
class CoIter(Stream):
    def __init__(self, make, update, state):
        self.make = make
        self.update = update
        self.state = state

    def head(self):
        return self.make(self.state)
    def tail(self):
        return CoIter(self.make, self.update, self.update(self.state))  
\end{lstlisting}
\end{contiguous}
For an example use of \lstinline`CoIter`, consider this class which
\lstinline`Maps` some \lstinline`trans`formation function over an existing
stream:
\begin{contiguous}
\begin{lstlisting}[language=Python]
class Maps(Stream):
    def __init__(self, stream, trans):
        self.stream = stream
        self.trans = trans

    def head(self): return self.trans(self.stream.head())
    def tail(self): return Maps(self.stream.tail(), self.trans)
\end{lstlisting}
\end{contiguous}
Each \lstinline`head` element is given by applying the
\lstinline`trans`formation to the \lstinline`head` of the current
\lstinline`stream`, while the \lstinline`tail` is computed by taking the
\lstinline`tail` of the underlying stream.  The underlying \lstinline`stream` is
changed on the recursive call to \lstinline`Maps`, so it must be part of its
evolving internal state, but the \lstinline`trans`formation stays the same.
Thus, the \lstinline`Maps` constructor of the above class definition is
equivalent to this application of \lstinline`Coiter`:
\begin{contiguous}
\begin{lstlisting}[language=Python]
def maps(s, f):
    return CoIter(
        lambda s: f(s.head()),
        lambda s: s.tail(),
        s)
\end{lstlisting}
\end{contiguous}

For example, we can use the stream of all \lstinline`nats` above to simulate the
stream of square numbers---\lstinline`0, 1, 4, 9, 16, ...`---like so:
\begin{lstlisting}[language=Python]
squares = maps(nats, lambda x: x*x)
\end{lstlisting}
As another example, here is the function which \lstinline`zips` together the
values of two streams:
\begin{contiguous}
\begin{lstlisting}[language=Python]
def zips(left, right, combine=lambda x, y: (x,y)):
    return CoIter(
        lambda state: combine(state[0].head(), state[1].head()),
        lambda state: (state[0].tail(), state[1].tail()),
        (left, right))
\end{lstlisting}
\end{contiguous}
By default, elements of the two streams \lstinline`x1, x2, x3, ...` and
\lstinline`y1, y2, y3, ...` are just \lstinline`combine`d as pairs:
\lstinline`(x1, y1), (x2, y2), (x3, y3), ...`.
% \begin{lstlisting}[language=Python]
% class Zips(Stream):
%     def __init__(self, left, right, combine=lambda x, y: (x,y)):
%         self.left = left
%         self.right = right
%         self.combine = combine

%     def head(self):
%         return self.combine(self.left.head(), self.right.head())

%     def tail(self):
%         return Zips(self.left.tail(), self.right.tail(), self.combine)
% \end{lstlisting}
This can be used to pair up the elements of a stream that appear sequentially
next to one another, allowing us to view them together by twos:
\begin{contiguous}
\begin{lstlisting}[language=Python]
def by_twos(stream, f=lambda x, y : (x, y)):
    return zips(stream, stream.tail(), f)
\end{lstlisting}
\end{contiguous}
For example, \lstinline`by_twos(nats)` simulates the stream
\lstinline`(0,1), (1,2), (2,3), (3,4), ...`
of all natural numbers paired with their successor.

\begin{remark}
\label{rm:ephemeral-iterators}

An experienced Python programmer might think ``these streams all look awfully
similar to iterators, why to just use them instead?''  The reason is that all
the \lstinline`Stream` objects above are \emph{persistent}: an object's
\lstinline`head` and \lstinline`tail` never changes, no matter how many times
these methods are called.  In contrast, iterators are \emph{ephemeral}: each
time the \lstinline`next` element of an iterator object is requested, the
response is different.

Fundamentally, the \lstinline`Iterator` interface---which only include the
single method \lstinline`next`---must execute it's task imperatively.  The
\lstinline`next` method returns the next element as stated, but it also
implicitly \emph{mutates} the iterator itself behind the scenes.  That way, the
next call to \lstinline`next` will return a new element, and not the current
one.  Instead, the \lstinline`Stream`s above separate this task into two
independent methods: \lstinline`head` only returns the first element and
\lstinline`tail` returns a new \lstinline`Stream` with the updated state.  As
such, \lstinline`head` can return the same element every time, and there is no
need to modify a \lstinline`Stream` object to behave like its \lstinline`tail`
on the next call.

The persistence allowed by the \lstinline`Stream` interface opens up new
possibilities.  For example, the \lstinline`by_twos` function above takes a
single stream and copies it, traversing the same stream \emph{twice}
simultaneously.  This operation does not make sense for an \lstinline`Iterator`,
which we can only traverse once.  Instead, the user of the \lstinline`Iterator`
must be responsible for remembering enough of the previous elements it returned
in order to simulate the persistent behavior required by \lstinline`by_twos`.
This could be done by embedding the logic of \lstinline`zips` directly into
\lstinline`by_twos` and explicitly keeping the next two elements at a time.  Or
it could be done generically like Python's standard \lstinline`tee` function,
which explicitly memoizes the elements across duplicate references to one
\lstinline`Iterator`.  Persistent \lstinline`Stream`s eliminate this
complication.
\end{remark}

We now know to co\-iterate in an object-oriented style, but what of
co\-recursion?  Previously we saw the functional version of co\-recursion that
could end the co\-recursive loop early (in Agda, \cref{sec:agda-corec}) and even
provide a continuation that could be resumed several times (in Scheme,
\cref{sec:scheme-corec}).  In either case, we needed to provide an alternative
exit path to provide the rest of the stream, rather than updating the internal
state.  A suitably object-oriented way to provide multiple exit paths is with
\emph{exceptions}.  For example, a Python class for \lstinline`CoRec`ursion can
be defined in terms of a \lstinline`StopCoRec` exception as follows:
\begin{contiguous}
\begin{lstlisting}[language=Python,basicstyle=\ttfamily\footnotesize]
class StopCoRec(Exception):
    def __init__(self, s):
        self.remainder = s

class CoRec(CoIter):
    def tail(self):
        try:
            return CoRec(self.make, self.update, self.update(self.state))
        except StopCoRec as done:
            return done.remainder  
\end{lstlisting}
\end{contiguous}
\lstinline`CoRec` is defined as a subclass of \lstinline`Coiter` because they
share many similarities.  Both contain three parts:
\begin{enumerate*}
\item a way to \lstinline`make` an element from the current state,
\item a way to \lstinline`update` the state, and
\item an initial value for the \lstinline`state`.
\end{enumerate*}
And in fact, the \lstinline`head` of both \lstinline`CoRec`ursion and
\lstinline`CoIter`ation is identical, so it does not need to be specified again
here.  However, the method for computing the \lstinline`tail` is different.
\lstinline`CoRec` expects that the \lstinline`update` function might raise an
exception, in which case it responds differently:
\begin{itemize}
\item If \lstinline`self.update(self.state)` returns normally, then the value 
  returned is used as the new state to continue co\-recursion.
\item Otherwise, \lstinline`self.update(self.state)` might raise a
  \lstinline`StopCoRec` exception.  In this case, co\-recursion ends and the
  remainder of the stream (contained within the exception) is returned as the
  tail of the \lstinline`CoRec`ursor.
\end{itemize}
For example, the \lstinline`scons` function can be defined via \lstinline`CoRec`
in Python as:
\begin{contiguous}
\begin{lstlisting}[language=Python]
def scons(hd, tl):
    def head(_): return hd
    def tail(_): raise StopCoRec(tl)
    return CoRec(head, tail, None)
\end{lstlisting}
\end{contiguous}
Note that the \lstinline`tail` of this \lstinline`CoRec`ursor will always stop
immediately, and just return the given stream \lstinline`tl`.

\begin{figure}
\centering
\begin{lstlisting}[language=Python]
class Ended(Exception): pass

class Ending(Stream):
    # head : self -> elem
    # tail : self -> Stream  throws  Ended

    def __iter__(self):
        curr = self
        while True:
            try:
                yield curr.head()
                curr = curr.tail()
            except Ended:
                break
\end{lstlisting}
\caption{An abstract class of \lstinline`Ending` streams in Python.}
\label{fig:python-ending-stream}
\end{figure}
Using exceptions to model alternate exits opens up a world of possibilities for
generalizing streams.  For example, we can define a new class for streams which
might eventually come to an end when their \lstinline`tail` raises an exception,
as shown in \cref{fig:python-ending-stream}.  With \lstinline`Ending` streams,
we need to expect that any call to \lstinline`tail` might raise the
\lstinline`Ended` exception, and handle it accordingly.  For example, the method
of iterating through the elements of the stream (\lstinline`__iter__`) needs to
be updated to account for this additional case: if the stream has
\lstinline`Ended`, then the iteration should stop.  The smallest
\lstinline`Ending` stream is the one that contains only a \lstinline`Single`
element:
\begin{contiguous}
\begin{lstlisting}[language=Python]
class Single(Ending):
    def __init__(self, only):
        self.value = only

    def head(self): return self.value
    def tail(self): raise Ended
\end{lstlisting}
\end{contiguous}
So that \lstinline`Single(1)` represents the stream which contains only the
value \lstinline`1`.  We can also \lstinline`append` one \lstinline`Ending`
stream onto another stream (which may or may not end) in terms of
\lstinline`CoRec` as:
\begin{contiguous}
\begin{lstlisting}[language=Python]
def append(prefix, suffix):
    def head(pre):    return pre.head()
    def tail(pre):
        try:          return pre.tail()
        except Ended: raise StopCoRec(suffix)

    return CoRec(head, tail, prefix)
\end{lstlisting}
\end{contiguous}
Similar to append modeled previously in Agda and Scheme, the state of the
\lstinline`CoRec`ursion is the \lstinline`prefix`.  However, since the
\lstinline`prefix` is now an \lstinline`Ending` stream containing at least one
element, there are fewer cases to consider: the \lstinline`head` is always the
\lstinline`head` of the remaining prefix, and the \lstinline`tail` depends on
whether or not that prefix has \lstinline`Ended`. If the prefix has more
elements, then the state is updated with its \lstinline`tail`.  But if the
prefix has \lstinline`Ended`, then the \lstinline`tail` of the
\lstinline`append` is just the suffix (dropping the single element remaining in
the prefix).

Co\-iteration is defined for \lstinline`Ending` streams the same as it is for
infinite ones, noting that \lstinline`Ended` exceptions are propagated
implicitly.  So we can convert any finite list into an \lstinline`Ending` stream
like so:
\begin{contiguous}
\begin{lstlisting}[language=Python]
class CoIterEnds(CoIter, Ending): pass

def stream_list(items):
    def increment(i):
        i += 1
        if i < len(items):
            return i
        else:
            raise Ended
        
    return CoIterEnds(lambda i: items[i], increment, 0)
\end{lstlisting}
\end{contiguous}
For example, we can represent the stream that counts down from \lstinline`3`,
\ie \lstinline`3, 2, 1, 0, 0, 0, ...`,
as the following application of \lstinline`append`, \lstinline`stream_list`, and
\lstinline`zeroes`:
\begin{lstlisting}[language=Python]
count_down = append(stream_list(range(3,0,-1)), zeroes)
\end{lstlisting}

\begin{figure}
\centering
\begin{lstlisting}[language=Python]
class Skipped(Exception):
    def __init__(self, skip=None):
        self.value = skip

class Skipping(Stream):
    # head : self -> elem  throws  Skipped
    # tail : self -> Stream

    def __iter__(self):
        curr = self
        while True:
            try:
                yield curr.head()
            except Skipped:
                pass
            curr = curr.tail()
\end{lstlisting}
\caption{An abstract class of \lstinline`Skiping` streams in Python.}
\label{fig:python-skipping-stream}
\end{figure}

In contrast to \lstinline`Ending` streams, we can also use exceptions to
represent
streams that can skip certain elements, as shown in
\cref{fig:python-skipping-stream}.  With \lstinline`Skipping` streams, we need
to expect that any request for the \lstinline`head` element might raise a
\lstinline`Skipped` exception, and handle it accordingly.  For example, the
method of iterating through elements of a \lstinline`Skipping` stream
(\lstinline`__iter__`) needs to be updated to account for this additional case:
if the current \lstinline`head` element is \lstinline`Skipped`, then iteration
must continue on through the remaining elements without \lstinline`yield`ing
anything.  Why might we want to skip elements explicitly?  Consider the task of
filtering a stream: removing the elements that do not pass some predicate.
Normally, filter is not an instance of \lstinline`CoIter` for infinite streams
because the result might not be another infinite stream.  For example, the
predicate \lstinline`lambda x: false` will reject every single element from a
\lstinline`stream`, so \lstinline`filter(stream, lambda x: false).head()` cannot
return anything.  However, filtering \emph{is} an instance of \lstinline`CoIter`
for \lstinline`Skipping` streams:
% \begin{lstlisting}[language=Python]
% class Filters(Skipping):
%     def __init__(self, stream, f):
%         self.stream = stream
%         self.check = f

%     def head(self):
%         x = self.stream.head()
%         if self.check(x):
%             return x
%         else:
%             raise Skipped(x)

%     def tail(self):
%         return Filters(self.stream.tail(), self.check)
% \end{lstlisting}
\begin{contiguous}
\begin{lstlisting}[language=Python]
class CoIterSkips(CoIter, Skipping): pass

def filters(stream, check):
    def head(s):
        x = s.head()
        if check(x):
            return x
        else:
            raise Skipped(x)

    return CoIterSkips(head, lambda s: s.tail(), stream)
\end{lstlisting}
\end{contiguous}
If the current element passes the \lstinline`check`, then it is returned
normally.  Otherwise, it is just \lstinline`Skipped`.  For example, we can
capture only the even square numbers as:
\begin{lstlisting}[language=Python]
even_squares = filters(squares, lambda x: x % 2 == 0)
\end{lstlisting}
In \lstinline`even_squares`, every other number will be \lstinline`Skipped`.  We
never have to worry about what the next element of the filtered stream is, or if
it will ever come.  So \lstinline`filter(stream, lambda x: false)` is
well-defined: it is just the infinite stream where every single element is
\lstinline`Skipped`.

%%% Local Variables:
%%% mode: latex
%%% TeX-master: "corec-prog"
%%% End:

\section{Java: Typed Interfaces for Corecursive Methods}
\label{sec:java-corec}

\lstset{language=Java}

While Python makes it easy to program with objects and exceptions, our
understanding of which methods might return certain exceptions, and which ones
do not, is completely informal.  This can be an issue for understanding code,
because the meaning of the different stream types (ones that might end or skip
elements) depends crucially on this implicit contract on when exceptions are
expected.

If we want a more formal description of the different stream interfaces, we can
instead look to a statically typed language like Java.  In particular, Java has
\emph{checked exceptions} which keep track of the exceptions that might be
thrown by a method in its type.  That way, we will statically know when calling
a \lstinline`tail` method might end, or when it is guaranteed to go on forever,
and the compiler keeps track of the difference for us.

However, before we delve into the different interfaces for streams, lets review
how exceptions interact with subtyping.  First, consider this functional
interface for a basic unary function from type \lstinline`A` to \lstinline`B`:
\begin{contiguous}
\begin{lstlisting}[language=Java]
public interface Function<A, B> {
    B apply(A arg);
}
\end{lstlisting}
\end{contiguous}
For example, Java 8 lets us write the lambda expression \lstinline`x -> x + x`,
which can be an object of type \lstinline`Function<Integer, Integer>`; its
\lstinline`apply` method will be defined as:
\begin{lstlisting}[language=Java]
  Integer apply(Integer x) { return x + x; }
\end{lstlisting}
However, an expression like \lstinline`x -> { throw new Exception(); }` cannot
have the same type \lstinline`Function<Integer, Integer>`.  Why not?  Because
the corresponding method definition
\begin{lstlisting}[language=Java]
  Integer apply(Integer x) { throw new Exception(); }
\end{lstlisting}
is \emph{not} well-typed.  This method throws an \lstinline`Exception`, and Java
requires this fact be included in the type of the method.

To allow for checked exceptions in functions, we have to include them explicitly
in the interface.  For example, we can generalize the \lstinline`Function`
interface above to this one, whose \lstinline`apply` method is stated to throw
an exception \lstinline`E`:
\begin{contiguous}
\begin{lstlisting}[language=Java]
public interface FunctionThrows<A, B, E extends Exception> {
    B apply(A arg) throws E;
}
\end{lstlisting}
\end{contiguous}
Now, both \lstinline`x -> x + x` and \lstinline`x -> { throw new Exception(); }`
can be given the same type
\lstinline`FunctionThrows<Integer,Integer,Exception>`.  That's because
\begin{contiguous}
\begin{lstlisting}[language=Java]
  Integer apply(Integer x) throws Exception {
    return x + x;
  }
  Integer apply(Integer x) throws Exception {
    throw new Exception();
  }
\end{lstlisting}
\end{contiguous}
are both well-typed.  Checked exceptions note which exceptions \emph{might} be
thrown by a method, but does not require them to.

Now that we have two similar functional interfaces, we can ask when can we use
objects of one type in place of the other.  Or in other words, which of
\lstinline`Function` or \lstinline`FunctionThrows` is a subtype of the other?
First, consider this method for explicitly converting a \lstinline`Function`
into a \lstinline`FunctionThrows`:
\begin{contiguous}
\begin{lstlisting}[language=Java]
public static<A, B, E extends Exception>
  FunctionThrows<A, B, E> neverThrows(Function<A, B> f) {
    return x -> f.apply(x);
}
\end{lstlisting}
\end{contiguous}
Is this well-typed?  Yes.  \lstinline`f.apply(x)` will never throw a checked
exception, but it doesn't matter that the new object \lstinline`x -> f.apply(x)`
doesn't throw an exception \lstinline`E`; checked exceptions are an allowance,
not a mandate.  In contrast, consider the reverse conversion:
\begin{contiguous}
\begin{lstlisting}[language=Java]
public static<A, B, E extends Exception>
  Function<A, B> mightThrow(FunctionThrows<A, B, E> f) {
    return x -> f.apply(x);
}
\end{lstlisting}
\end{contiguous}
Is this well-typed?  No.  Here, \lstinline`f.apply(x)` might throw an exception
\lstinline`E`.  This is forbidden by the \lstinline`Function` interface, whose
\lstinline`apply` method cannot throw \emph{any} checked exception.  Thus,
\lstinline`Function<A, B>` can be seen as a subtype of
\lstinline`FunctionThrows<A, B, E>` (for any \lstinline`E`), but not vice versa.

With this in mind, let's now consider the different interfaces for (possibly)
infinite streams.  Previously in \cref{sec:python-corec}, we saw
\lstinline`Streams` that expected to be infinite, with \lstinline`head` and
\lstinline`tail` methods that always return.  That was generalized to subclasses
of streams that might come to an end (when \lstinline`tail` raises an exception)
or that might skip certain elements (when \lstinline`head` raises an exception).
But, if we look at the above subtyping relationship between methods that might
raise exceptions versus ones that don't, we see that the
\cref{sec:python-corec}'s subclasses of streams are backwards!  The interface of
truly infinite streams---whose methods \emph{never} raise an exception---should
be a subtype of both ending and skipping streams---whose \lstinline`tail` might
have \lstinline`Ended` or \lstinline`head` might be \lstinline`Skipped`---not
the other way around.

\begin{figure}
\begin{subfigure}[t]{\textwidth}
\begin{lstlisting}[language=Java]
public interface Stream<A> {
    public A head() throws Skipped;
    public Stream<A> tail() throws Ended;

    public class Ended extends Exception { public Ended() {} }
    public class Skipped extends Exception { public Skipped() {} }
}  
\end{lstlisting}
\caption{General \lstinline`Stream`s that might skip or end.}
\label{fig:java-general-stream}
\end{subfigure}
\\[1em]
\begin{subfigure}[t]{\textwidth}
\begin{lstlisting}[language=Java]
public interface Ending<A> extends Stream<A> {
    public A head();
    public Ending<A> tail() throws Ended;
}
\end{lstlisting}
\caption{\lstinline`Ending` streams, that never skip elements.}
\label{fig:java-ending-stream}
\end{subfigure}
\\[1em]
\begin{subfigure}[t]{\textwidth}
\begin{lstlisting}[language=Java]
public interface Skipping<A> extends Stream<A> {
    public A head() throws Skipped;
    public Skipping<A> tail();
}
\end{lstlisting}
\caption{\lstinline`Skipping` streams, that never end.}
\label{fig:java-skipping-stream}
\end{subfigure}
\\[1em]
\begin{subfigure}[t]{\textwidth}
\begin{lstlisting}[language=Java]
public interface Infinite<A> extends Ending<A>, Skipping<A> {
    public A head();
    public Infinite<A> tail();
}
\end{lstlisting}
\caption{\lstinline`Infinite` streams that never skip or end.}
\label{fig:java-infinite-stream}
\end{subfigure}

\caption{Four different \lstinline`Stream` interfaces in Java.}
\label{fig:java-streams}
\end{figure}

The four different interfaces of streams given in \cref{fig:java-streams}
present all these possible combinations of checked exceptions for ending and
skipping, and take advantage of the subtyping relationship between them.  The
largest super-interface of \lstinline`Streams` (\cref{fig:java-general-stream})
that could either skip or end is given first.  Then two sub-interfaces refine
this one: \lstinline`Ending` streams (\cref{fig:java-ending-stream}) never skip
elements and \lstinline`Skipping` streams (\cref{fig:java-skipping-stream})
never end.  Finally, truly \lstinline`Infinite` streams
(\cref{fig:java-infinite-stream}) follow a sub-interface combining \emph{both}
of these refinements together.  Take note that the types ensure that refinements
made by each sub-interface are \emph{hereditary} among different stream
interfaces.  For example, an \lstinline`Infinite` stream is guaranteed to have a
\lstinline`head` and \lstinline`tail` now, and also \emph{all} of its
\lstinline`tails` share this promise because the \lstinline`tail` of an
\lstinline`Infinite` stream is another \lstinline`Infinite` stream.  Similar
heredity follows for the weaker promise of \lstinline`Ending` and
\lstinline`Skipping` streams.  Alternatively, we could have said that the
\lstinline`tail` method of each interface returns any \lstinline`Stream`, but
this would not express the intent of the hereditary promise.  If the
\lstinline`tail` of an \lstinline`Infinite` stream were just any
\lstinline`Stream`, then its \lstinline`tail` would loose the promise that
future \lstinline`head` or \lstinline`tail` methods definitely return.

Unlike an \lstinline`Ending` stream---which must contain at least one
element---a general \lstinline`Stream` may truly be empty, as per the following
class:
\begin{contiguous}
\begin{lstlisting}[language=Java]
public class Empty<A> implements Stream<A> {
    public Empty() { }
    public A head() throws Skipped { throw new Skipped(); }
    public Stream<A> tail() throws Ended { throw new Ended(); }
}
\end{lstlisting}
\end{contiguous}
Like \lstinline`Single` (from \cref{sec:python-corec}) and \lstinline`Empty`
object has no \lstinline`tail`, but unlike \lstinline`Single(x)` it \emph{also}
has no \lstinline`head`.  A \lstinline`Skipping` stream can simulate emptiness
by always skipping its \lstinline`head` element:
\begin{contiguous}
\begin{lstlisting}[language=Java]
public class AlwaysSkips<A> implements Skipping<A> {
    public AlwaysSkips() { }
    public A head() throws Skipped { throw new Skipped(); }
    public Skipping<A> tail() { return this; }
}
\end{lstlisting}
\end{contiguous}
In contrast with \lstinline`Empty`---whose observer can discover quite quickly
that it will never produce anything---an outside observer will not be able to
tell (in any finite amount of time) whether or not \lstinline`AlwaysSkips` is
effectively empty.  The observer will find that every \lstinline`tail` will
skip, but it can never be sure whether or not an element will eventually be
found further in.  This issue hints at a problem with \lstinline`Skipping`
streams (and \lstinline`Streams` in general).

For example, consider the \lstinline`take` method from \cref{sec:python-corec},
which returns a finite list of the first \lstinline`n` elements of a stream.
What would
\lstinline`new AlwaysSkips<A>().take(1)`
do?  It would loop forever, forever looking for that first non-skipped element
of the stream, that will never come.  In other words, operations like
\lstinline`take`, \lstinline`drop`, or sequential iteration through the elements
of a stream are \emph{unsafe} for \lstinline`Skipping` streams (let alone
general \lstinline`Streams`), because they might loop forever.  The proper place
to introduce them in the interface hierarchy of \cref{fig:java-streams} is in
\lstinline`Ending` streams, which can be given as these default implementations
in terms of \lstinline`head` and \lstinline`tail` that could be added to the
interface in \cref{fig:java-ending-stream}:
\begin{contiguous}
\begin{lstlisting}[language=Java]
default public Ending<A> drop(int n) throws Ended {
    Ending<A> dropped = this;
    for (int i = 0; i < n; i++) {
        dropped = dropped.tail();
    }
    return dropped;
}
\end{lstlisting}
\end{contiguous}
\begin{contiguous}
\begin{lstlisting}[language=Java]
default public Vector<A> take(int n) {
    Vector<A> taken = new Vector<A>(n);
    Ending<A> dropped = this;
    for (int i = 0; i < n; i++) {
        taken.add(dropped.head());
        try {
            dropped = dropped.tail();
        } catch (Ended e) {
            break;
        }
    }
    return taken;
}
\end{lstlisting}
\end{contiguous}
Through inheritance, we can \lstinline`take` and \lstinline`drop` elements from
\lstinline`Infinite` streams, too.%
\footnote{\lstset{basicstyle=\ttfamily\footnotesize} Although, the
  \lstinline`Infinite` interface can override the \lstinline`drop` method to
  provide a tighter type, since \lstinline`drop`ping elements from an
  \lstinline`Infinite` stream will never be \lstinline`Ended` prematurely, and
  will return another \lstinline`Infinite` stream.}
Attempting to \lstinline`take` or \lstinline`drop` some elements from an
\lstinline`Ending` stream might come up short, but they will never loop forever
looking for more.

Instead, if we really insist on taking or iterating through the elements of a
\lstinline`Skipping` stream, the common, risky component is fast forwarding past
all the skipped elements.  This is the fundamentally unsafe operation, which
could easily loop forever, and is performed by the following class for
generating \lstinline`Infinite` streams from \lstinline`Skipping` ones:
\begin{contiguous}
\begin{lstlisting}[language=Java]
public class FastForward<A> implements Infinite<A> {
    private boolean compressed;
    private Skipping<A> skips;
\end{lstlisting}
\end{contiguous}
\begin{contiguous}
\begin{lstlisting}[language=Java]
    public FastForward(Skipping<A> s) {
        this.skips = s;
        this.compressed = false;
    }
\end{lstlisting}
\end{contiguous}
\begin{contiguous}
\begin{lstlisting}[language=Java]
    public A head() {
        while (true) {
            try {
                A hd = this.skips.head();
                this.compressed = true;
                return hd;
            } catch (Skipped e) {
                this.skips = this.skips.tail();
            }
        }
    }
\end{lstlisting}
\end{contiguous}
\begin{contiguous}
\begin{lstlisting}[language=Java]
    public FastForward<A> tail() {
        if (!this.compressed) {
            this.head();
        }
        return new FastForward<A>(this.skips.tail());
    }
}
\end{lstlisting}
\end{contiguous}
The first time the \lstinline`head` element is requested, then the underlying
\lstinline`Skipping` stream is queried until it finally returns a
\lstinline`head` element (or loops forever trying).  That point in the stream is
remembered for future calls to \lstinline`head`, to find it in constant time.
Computing the \lstinline`tail` requires that we have at least found the next
\lstinline`head` element, so that we can move past it.  In a stream like
\lstinline`Skipped, Skipped, 0, Skipped, 1, 2, Skipped, 3, ...`
we do not want to count the \lstinline`Skipped` elements, so its tail should at
least be
\lstinline`Skipped, 1, 2, Skipped, 3, ...`

So what are the pattern for the well-founded ways of generating streams, and how
do they differ between the four interfaces?  The definition of co\-iteration for
\lstinline`Infinite` streams in Java resembles all the previous co\-iterators,
and especially the one in Python:
\begin{contiguous}
\begin{lstlisting}[language=Java,basicstyle=\ttfamily\footnotesize]
public class InfiniteCoIter<B,A> implements Infinite<A> {
    private final B state;
    private final Function<B,A> make;
    private final Function<B,B> update;
\end{lstlisting}
\end{contiguous}
\begin{contiguous}
\begin{lstlisting}[language=Java,basicstyle=\ttfamily\footnotesize]
    public InfiniteCoIter(B x, Function<B,A> f, Function<B,B> g) {
        this.state = x;
        this.make = f;
        this.update = g;
    }
\end{lstlisting}
\end{contiguous}
\begin{contiguous}
\begin{lstlisting}[language=Java,basicstyle=\ttfamily\footnotesize]
    public A head() {
        return this.make.apply(this.state);
    }
\end{lstlisting}
\end{contiguous}
\begin{contiguous}
\begin{lstlisting}[language=Java,basicstyle=\ttfamily\footnotesize]
    public CoIterInfinite<B,A> tail() {
        B next = this.update.apply(this.state);
        return new InfiniteCoIter<B,A>
            (next, this.make, this.update);
    }
}
\end{lstlisting}
\end{contiguous}
But this only lets us generate \lstinline`Infinite` streams.  What if we want to
generate a general \lstinline`Stream` that might skip or end?  Perhaps
surprisingly, the only difference is a change to the types of functional
parameters that are allowed.  In particular, the function that describes the
base case---that \lstinline`make`s the current element from the
\lstinline`state`---is allowed to throw a \lstinline`Skipped` exception,
effectively skipping that state.  The function that describes the co\-inductive
case---that \lstinline`update`s the \lstinline`state`---is allowed to throw the
\lstinline`Ended` exception, to end co\-iteration and the stream.  Other than
this change of types, the code is identical (modulo co\-recursively calling the
\lstinline`StreamCoIter` constructor rather than \lstinline`InfiniteCoIter`):%
\footnote{The in-between versions of using co\-iteration to generate
  \lstinline`Ending` and \lstinline`Skipping` streams are defined by allowing
  only \lstinline`update` to throw \lstinline`Ended` or allowing only
  \lstinline`make` to throw \lstinline`Skipped`, respectively.}
\begin{contiguous}
\begin{lstlisting}[language=Java,basicstyle=\ttfamily\footnotesize]
public class StreamCoIter<B,A> implements Stream<A> {
    private final B state;
    private final FunctionThrows<B,A,Skipped> make;
    private final FunctionThrows<B,B,Ended> update;
\end{lstlisting}
\end{contiguous}
\begin{contiguous}
\begin{lstlisting}[language=Java]
    public StreamCoIter(B x,
                        FunctionThrows<B,A,Skipped> f,
                        FunctionThrows<B,B,Ended> g) {
        ...
    }
    public A head() throws Skipped { ... }
    public StreamCoIter<B,A> tail() throws Ended { ... }
}
\end{lstlisting}
\end{contiguous}
For example, we know that \lstinline`map`ping over an \lstinline`Infinite`
stream is an instance of co\-iteration; in Java the instantiation looks like
this:
\begin{contiguous}
\begin{lstlisting}[language=Java,basicstyle=\ttfamily\footnotesize]
public static<A,B> Infinite<B> map(Infinite<A> stream,
                                   Function<A,B> trans) {
    return new InfiniteCoIter<Infinite<A>, B>
        (stream,
         s -> trans.apply(s.head()),
         s -> s.tail());
}
\end{lstlisting}
\end{contiguous}
Instead, when given a general \lstinline`Stream` that might skip elements, the
\lstinline`trans`formation function might want to skip elements, too. This gives
us the following more general operation that sometimes maps over the elements of
a \lstinline`Stream`, and sometimes skips them:
\begin{contiguous}
\begin{lstlisting}[language=Java,basicstyle=\ttfamily\footnotesize]
public static<A,B>
    Stream<B> mapSometimes(Stream<A> stream,
                           FunctionThrows<A,B,Skipped> trans) {
    return new StreamCoIter<Stream<A>, B>
        (stream,
         s -> trans.apply(s.head()),
         s -> s.tail());
}
\end{lstlisting}
\end{contiguous}
Again, note that the body of this function is nearly identical (up to a change
of the co\-iteration constructor).  What's different is that either applying the
\lstinline`trans`formation function or asking for the \lstinline`head` of
\lstinline`s` might implicitly throw a \lstinline`Skipped` exception.  Likewise,
\lstinline`s.tail()` might implicitly come to an \lstinline`Ended` stream.
Because \lstinline`mapSometimes` can accept more function parameters, it
encompasses the usual \lstinline`filter` function.  In particular,
\lstinline`filter` just uses a \lstinline`boolean` predicate to
\lstinline`check` each element, and the ones that fail are \lstinline`Skipped`:
\begin{contiguous}
\begin{lstlisting}[language=Java,basicstyle=\ttfamily\footnotesize]
public static<A>
    Stream<A> filter(Stream<A> stream, Function<A, Boolean> check) {
    return mapSomtimes
        (stream,
         x -> {
            if (check.apply(x)) {
                return x;
            } else {
                throw new Skipped();
            }
        });
}
\end{lstlisting}
\end{contiguous}

This covers co\-iteration, but what of co\-recursion, which is able to more
efficiently perform operations like appending a prefix on top of a stream?  Here
we finally come to the point where Java's type system restricts us.  In Python,
we used an exception, \lstinline`HaltCoRec` to capture and effectively return
the remainder of the stream.  In Java, we cannot catch generic exceptions, and
so we have to pick up-front a concrete type of streams---say, streams of
integers---that will be given if co\-recursion halts early.  Still, we can give
the following class for co\-recursively generating a
\lstinline`Stream<Integer>`:
\begin{contiguous}
\begin{lstlisting}[language=Java,basicstyle=\ttfamily\footnotesize]
public class HaltCoRec extends Exception {
    private final Stream<Integer> rest;
    public HaltCoRec(Stream<Integer> s) { this.rest = s; }
    public Stream<Integer> remainder() { return this.rest; }
}
\end{lstlisting}
\end{contiguous}
\begin{contiguous}
\begin{lstlisting}[language=Java]
public class CoRec<B> implements Stream<Integer> {
    private final B state;
    private final FunctionThrows<B, Integer, Skipped> make;
    private final FunctionThrows<B, B, HaltCoRec> update;
\end{lstlisting}
\end{contiguous}
\begin{contiguous}
\begin{lstlisting}[language=Java]
    public CoRec(B x,
                 FunctionThrows<B, Integer, Skipped> f,
                 FunctionThrows<B, B, HaltCoRec> g) {
        this.state = x;
        this.make = f;
        this.update = g;
    }
\end{lstlisting}
\end{contiguous}
\begin{contiguous}
\begin{lstlisting}[language=Java]
    public Integer head() throws Skipped {
        return this.make.apply(this.state);
    }
\end{lstlisting}
\end{contiguous}
\begin{contiguous}
\begin{lstlisting}[language=Java]
    public Stream<Integer> tail() {
        try {
            B next = this.update.apply(this.state);
            return new CoRec<B>(next, this.make, this.update);
        } catch (HaltCoRec halt) {
            return halt.remainder();
        }
    }
}
\end{lstlisting}
\end{contiguous}
Note that here, the \lstinline`make` operation might skip the current element,
like in the co\-iterator for \lstinline`Streams`.  However, instead of ending
the stream entirely, the \lstinline`update` operation might throw the
\lstinline`HaltCoRec` exception which contains the \lstinline`remainder` of the
\lstinline`Stream`.  Co\-recursors for the other types of
streams---\lstinline`Infinite`, \lstinline`Ending`, and \lstinline`Skipping`
ones---can be obtained by changing the type of the \lstinline`remainder`
contained in the \lstinline`HaltCoRec` exception, and by preventing or allowing
\lstinline`Skipped` exceptions in \lstinline`make` as appropriate.

%%% Local Variables:
%%% mode: latex
%%% TeX-master: "corec-prog"
%%% End:

\section{Conclusion}
\label{sec:conclusion}

Surely, structural co\-recursion is especially elegant in purely functional
languages for \emph{unfolding} lazy data structures.  Yet, we have found that
this same core idea can also be found in many other contexts, too.  Rather than
laziness and purity, we base our shared notion of structural co\-recursion on
\emph{codata} \cite{CodataInAction}, which appears in various guises in a wider
variety of programming languages.  This way, we are able to show how write
programs with structural co\-recursion in strict languages.  This new way of
thinking empowers us to combine structural co\-recursion with computational
effects---like first-class control---which increases the combined expressive
power and allows us brand new algorithms that aren't possible in a pure
language.  We can even leave the functional paradigm entirely to rephrase
structural co\-recursion in terms of concepts familiar in common object-oriented
languages.

Allowing the paradigm shift suggests that some of the iconic techniques that are
unique to functional languages may be applicable within object-oriented
programs, too.  For example, we are in the process of translating all the
applications of ``why functional programming matters'' \cite{hughes89} to
commonly-used object-oriented languages.  We believe that the ideas of
functional programming are truly universal, and can and should be employed by
the broader audience of \emph{all} programmers, even the non-functional ones.

%%% Local Variables:
%%% mode: latex
%%% TeX-master: "corec-prog"
%%% End:

\section*{Acknowledgments}
\noindent
This work is supported by the National Science Foundation under Grant
No.~1719158.

\bibliographystyle{JFPlike}
\bibliography{corec-prog}

\label{lastpage01}

\end{document}